\documentclass[aps,prd,superscriptaddress,nofootinbib,eqsecnum,twocolumn]{revtex4}

\pdfoutput=1

\usepackage{amsfonts}
\usepackage{amsmath}
\usepackage{amssymb}
\usepackage{graphicx,color}
\usepackage{float}
\usepackage{hyperref}
\usepackage{subfigure}
\usepackage{mathtools,slashed}
\usepackage{ulem}
\usepackage{lipsum}
 
\def\be {\begin{equation}}
\def\ee {\end{equation}}
\def\bea {\begin{eqnarray}}
\def\eea {\end{eqnarray}}
\def\bc {\begin{center}}
\def\ec {\end{center}}

\def\mn {\mu\nu}
\def\mr {\mu\rho}
\def\rn {\rho\nu}
\def\nn {\nonumber}
\newcommand \Tr{\operatorname{\text{Tr}}}
\def\sp {\shortparallel}

\date{\today}

\begin{document}
\title{Heavy quark dynamics in a strongly magnetized quark-gluon plasma} 
\author{Aritra Bandyopadhyay}\email{aritrabanerjee.444@gmail.com}
 \affiliation{Guangdong Provincial Key Laboratory of Nuclear Science, Institute of Quantum Matter, South China Normal University, Guangzhou 510006, China}
 \affiliation{Guangdong-Hong Kong Joint Laboratory of Quantum Matter, Southern Nuclear Science Computing Center, South China Normal University, Guangzhou 510006, China}
 \affiliation{Physics Department and Center for Exploration of Energy and Matter, Indiana University, 2401 N Milo B. Sampson Lane, Bloomington, IN 47408, USA}
 
\author{Jinfeng Liao} \email{liaoji@indiana.edu}
 \affiliation{Physics Department and Center for Exploration of Energy and Matter, Indiana University, 2401 N Milo B. Sampson Lane, Bloomington, IN 47408, USA}

\author{Hongxi Xing}\email{hxing@m.scnu.edu.cn}
  \affiliation{Guangdong Provincial Key Laboratory of Nuclear Science, Institute of Quantum Matter, South China Normal University, Guangzhou 510006, China}
 \affiliation{Guangdong-Hong Kong Joint Laboratory of Quantum Matter, Southern Nuclear Science Computing Center, South China Normal University, Guangzhou 510006, China}

\begin{abstract}{
We present a calculation of the heavy quark momentum diffusion coefficients in a quark-gluon plasma under the presence of a strong external magnetic field, within the Lowest Landau Level (LLL) approximation.  In particular, we apply the Hard Thermal Loop (HTL) technique for the resummed effective gluon propagator, generalized for a hot and magnetized medium. Using the derived effective HTL gluon propagator and the LLL quark propagator we analytically derive the full results for the longitudinal and transverse momentum diffusion coefficients for charm and bottom quarks beyond the static limit. We also show numerical results for these coefficients in  two special cases where the heavy quark is moving either parallel or perpendicular to the magnetic field.}
\end{abstract}

\maketitle
\newpage

\section{Introduction}

It is well-known that some stellar objects (e.g. neutron stars, anomalous X-ray pulsars), where nuclear matter are assumed to be under extreme conditions, possess large surface magnetic fields~\cite{Chakrabarty:1997ef}. Such strong fields are also found to be present in non-central heavy ion collisions (HIC), sourced by the fast-moving and positively-charged protons of the colliding nuclei. Sophisticated numerical simulations have demonstrated that the initial strength of this magnetic field can be very high, $eB\sim \hat{O}(1) m_\pi^2$ at RHIC and $eB\sim \hat{O}(10)m_\pi^2$ at LHC \cite{Skokov:2009qp,Deng:2012pc,Bloczynski:2012en,Tuchin:2014iua,bzdak,McLerran}, and that on average it points in the direction perpendicular to the reaction plane. 
 

The presence of the strong and anisotropic magnetic field in the non-central HICs could potentially induce observable effects in these collisions.  For example, the magnetic field could lead to novel transport phenomena such as the chiral magnetic effect~\cite{cme1,cme2,cme3}, chiral magnetic wave~\cite{Burnier:2011bf} as well as charge-dependent directed flow~\cite{Gursoy:2014aka,Gursoy:2018yai,Das:2016cwd,Dubla:2020bdz}. The influence of strong magnetic fields on the photon and dilepton productions from quark-gluon plasma have also been studied extensively~\cite{Basar:2012bp,Ayala:2016lvs,Wang:2020dsr,Tuchin:2013bda,Sadooghi:2016jyf,Bandyopadhyay:2016fyd,Bandyopadhyay:2017raf,Ghosh:2018xhh,Islam:2018sog,Das:2019nzv}, which may  possibly help explain the observed large anisotropy of photon emissions by PHENIX~\cite{phenix}.   Such a strong magnetic field, introducing an extra scale in the quark-gluon plasma (QGP) in addition to the usual temperature and chemical potential, has also triggered significant interest in theoretically understanding the phase structures and properties of a strongly magnetized medium. For example, there have been a lot of studies on the finite temperature magnetic catalysis (MC)~\cite{mcat1,mcat2,mcat3}, the  inverse magnetic catalysis (IMC)~\cite{Bali,Farias:2014eca,Farias:2016gmy,Mueller:2015fka,Ayala:2014iba,Ayala:2014gwa,Ayala:2015bgv}, as well as other thermodynamic properties~\cite{Ding:2020hxw,Ding:2021cwv}.  For various developments along these directions, see recent reviews in e.g.~\cite{Kharzeev:2012ph,Shovkovy,Elia,Fukushima,Mueller, Miransky,Kharzeev:2015znc,Kharzeev:2020jxw,Fukushima:2018grm,Li:2020dwr,Liu:2020ymh,Gao:2020vbh,Bandyopadhyay:2020zte,Andersen:2014xxa,Andersen:2021lnk}.



The dynamical evolution of heavy quarks (HQ) serves as an important probe for the properties of strongly interacting hot quark-gluon plasma created in heavy ion collisions. Because of their large mass compared to the temperature scale, HQs are generated at the early stage of the initial hard scatterings  and are ``external'' to the bulk thermal medium. These heavy quarks traverse through the fireball and experience drag forces as well as random ``kicks''  from  the thermal partons in the bulk medium. A widely adopted approach to describe such HQ dynamics is  to use the Langevin equations for describing HQ in-medium evolution. The essential theoretical inputs needed for this approach include the HQ momentum drag and diffusion coefficients. These parameters are known to sensitively influence the phenomenological modelings of HQ dynamics and the predictions for experimental observables~\cite{Rapp:2018qla}. Many efforts have been made to compute these HQ transport coefficients in the quark-gluon plasma. A number of results were obtained when the heavy quarks are considered to be static with its much heavier mass as the highest scale of the system~\cite{CaronHuot:2007gq,CaronHuot:2008uh,Singh:2018wps}, known as the static limit of the HQ.  These computations typically employ the   Hard Thermal Loop (HTL) resummation method for the hot medium~\cite{Braaten:1991jj,Braaten:1991we,Thoma:1990fm,Moore:2004tg,Beraudo:2009pe,Monteno:2011gq}.   
Though it is easier to work within the static limit, which is  a valid approximation for low-momentum charm and bottom quarks, there is the strong need for going beyond the static limit, given that current HIC measurements for heavy flavor sector extend well into high momentum region where the transverse momentum scale could be much larger than the charm or  bottom quark masses.

The presence of strong magnetic field brings interesting new questions about HQ dynamics, namely the magnetic field effect on the HQ transport coefficients in a highly magnetized quark-gluon plasma.  There have been some recent developments on the HQ dynamics both within and beyond the static limit ~\cite{Fukushima:2015wck,Sadofyev:2015tmb,Kurian:2019nna,Singh:2020faa,Singh:2020fsj}, also within the holographic approach~\cite{Finazzo:2016mhm}. Most of those calculations consider the Lowest-Landau-Level (LLL) approximation, which for a thermal medium suggests the regime $eB \gg T^2$. On top of that, the HQ mass ($M$) is assumed to be the largest scale of the system, resulting in the scale hierarchy $M\gg \sqrt{eB} \gg T$. Similar to Ref~\cite{Fukushima:2015wck}, here we also work within a further constraint $\alpha_s eB \ll T^2$, $\alpha_s$ being the strong coupling, such that one can neglect the soft self energy corrections of the LLL quarks and gluons while evaluating the scattering rate.  
The presence of an  external magnetic field pointing at a fixed direction also breaks isotropy of the system, therefore even within the static limit of HQ, there will be two momentum diffusion coefficients, i.e. in the longitudinal and transverse directions of the magnetic field. Going beyond the static limit, there will be nontrivial interplay between the magnetic field direction and the HQ momentum direction, making the problem even more complex and challenging. Clearly, a lot more needs to be understood for HQ transport coefficients in a magnetized quark-gluon plasma.  

In this paper, we aim to address this important problem, namely the  calculation of the heavy quark momentum diffusion coefficients beyond the static limit in a quark-gluon plasma under the presence of a strong external magnetic field.   
Considering a HQ moving with a velocity $\vec{v}$ in presence of an anisotropic $\vec{B} = B \hat{z}$, 
 we analytically derive the full results for the longitudinal and transverse momentum diffusion coefficients  for charm and bottom quarks. 
 We will adopt the the Lowest Landau Level (LLL) approximation for medium quark propagators in the regime $M\gg \sqrt{eB} \gg T$ and use the HTL technique  for the resummed effective gluon propagators generalized for a hot and magnetized medium.  
 We also show numerical results for these coefficients in  two special cases where the heavy quark is moving either parallel or perpendicular to the external magnetic field ($\vec{v} \sp \vec{B}$ and $\vec{v} \perp \vec{B}$). 
 
 The rest of this paper is organized as follows. In section \ref{sec2} we discuss the basic formalism required to study the HQ dynamics, both for $B=0$ and $B \neq 0$,  within and beyond the static limit. In the following section (section \ref{sec3}) we compute the scattering rate for both $B=0$ and $B\neq 0$ beyond the static limit. In section \ref{sec4} we evaluate the final expressions for   the momentum diffusion coefficients of HQ in a strongly magnetized medium for both $\vec{v} \sp \vec{B}$ and $\vec{v}\perp\vec{B}$. Section \ref{sec5} contains our results and corresponding discussions. Finally we summarize and conclude in section \ref{sec6}.


\section{Formalism}
\label{sec2}
In the present work we focus on the HQ dynamics, where the HQ is assumed to be relativistic (i.e. beyond the static limit) in presence of a hot and magnetized medium. We will start the current section by discussing the $B=0$ case and gradually move in to the $B\neq 0$ cases, within and beyond the static limit. 

\subsection{HQ dynamics without magnetic field}

In absence of the external magnetic field, there is only one external scale from heavy quarks, i.e. $M \gg T$. Because of the fact that it takes many collisions to substantially change the momentum of the HQ, the interaction of the HQ with the medium can be approximated as uncorrelated momentum kicks. 
The corresponding dynamics follows the Langevin equation as
\begin{equation}
\frac{dp_i}{dt} = \xi_i(t) - \eta_D p_i, ~\langle \xi_i(t)\xi_j(t^\prime)\rangle = \kappa\delta_{ij}\delta(t-t^\prime),
\label{langevin1}
\end{equation}
where $(i,j)=(x,y,z)$ and $\xi_i(t)$ represents the uncorrelated momentum kicks. $\eta_D$ and $\kappa$ are respectively known as the momentum drag and diffusion coefficient in the static limit (i.e. with punishingly small $p$). Assuming $t>\eta_D^{-1}$, the solution of the above differential equation can be given as 
\begin{equation}
p_i(t) = \int\limits_{-\infty}^t dt^\prime e^{\eta_D(t^\prime-t)} 
\xi_i(t^\prime). 
\end{equation} 
As a result of the random kicks from medium particles, the HQ  momentum broadening (as quantified by the mean squared value of $p$) changes at a rate of 
\begin{equation}
 \frac{d}{dt}\langle p^2 \rangle = 3 \kappa 
\end{equation}
where $3\kappa$ is the momentum diffusion rate (i.e. mean squared momentum transfer per unit time) with the factor 3 coming from the 3 isotropic spatial dimensions. The coefficients $\kappa$ and $\eta_D$ are connected via the well-known fluctuation-dissipation relation. 

However, in high energy collisions, the charm and bottom quark spectra suggest a finite transverse momentum in general. Hence the relativistic case becomes  important to study. For this case, we consider HQ with finite velocity $\gamma v \lesssim 1$. In this kinematic regime, $p=\gamma M v \sim M$, i.e. the HQ momentum and mass are of similar scale. 
Now, considering the HQ is moving in a particular direction, we have the generalized Langevin equation as: 
\begin{subequations}
\bea
\frac{dp_i}{dt} = \xi_i(t) - \eta_D(p) p_i, \\
\langle \xi_i(t)\xi_j(t^\prime)\rangle = \kappa_{ij}(\vec p)\delta(t-t^\prime),
\label{langevin2}
\eea
\end{subequations}
where 
\bea
\kappa_{ij}({\vec{p}}) = \kappa_L(p)~ \hat{p}_i\hat{p}_j + \kappa_T(p) \left( \delta_{ij}-\hat{p}_i\hat{p}_j\right),
\eea
where $\hat{p}_i$ is the HQ momentum unit vector along specific direction $i$ with $(i,j) = (x,y,z)$. $\kappa_L$ and $\kappa_T$ are the longitudinal and transverse momentum diffusion coefficients respectively. Compared with the static case we can see that the anisotropy generated from the movement of HQ in a preferred direction  breaks down the  $\kappa$ into longitudinal and transverse parts, i.e. $3\kappa \to  \kappa_L + 2\kappa_T$. 
These anisotropic coefficients quantify the momentum diffusion rate due to scatterings with medium particles in the directions parallel or perpendicular to the HQ momentum: 
\begin{subequations}
\bea
\frac{1}{2}\frac{d}{dt} \langle (\Delta p_T)^2\rangle &\equiv& \kappa_T(p), \\
\frac{d}{dt} \langle (\Delta p_L)^2\rangle &\equiv& \kappa_L(p),
\eea
\end{subequations}
with $p_L$ and $p_T$ representing longitudinal and transverse momentum components. 
Note that since the $\eta_D$ becomes momentum-dependent, the relevant time scale set by $\sim 1/\eta_D$ would also become momentum-dependent. Nevertheless for the kinetic regime we consider (with $\gamma v \lesssim 1$), the HQ mass and HQ momentum are of similar scale and it is plausible to expect that the $1/\eta_D$ would remain at the same order of magnitude for the momentum regime of our interest. 

 The uncorrelated momentum kicks in a finite temperature medium originate from  the scattering processes of thermally populated light quarks and gluons with the heavy quark, i.e. $2\leftrightarrow 2$ scattering processes $qH\rightarrow qH$ and $gH\rightarrow gH$ ($q \rightarrow$ quark, $g\rightarrow$ gluon and $H\rightarrow$ HQ). At leading order in strong coupling, these scatterings are mediated by one-gluon exchange (see Fig.~\ref{hq_sqme_alt}), and the scattering particles can be considered as quasiparticles in thermally equilibrated matter. In the rest frame of the plasma, the Compton scattering is suppressed by the scale $T/M$ and hence both the $qH\rightarrow qH$ and $gH\rightarrow gH$ processes predominantly occur via the $t$-channel gluon exchange. Hence the momentum broadening rates  i.e. $\kappa_L$ and  $\kappa_T$ can be  directly expressed through   the scattering rate $\Gamma$ of the t-channel gluon exchange, as follows: 
\begin{subequations}
\bea
\kappa_L &=& \int d^3q~\frac{d\Gamma}{d^3q} ~q_L^2, \\
\kappa_T &=& \frac{1}{2}\int d^3q ~\frac{d\Gamma}{d^3q} ~q_T^2.
\eea 
\end{subequations}
Again the corresponding drag coefficients can be related to the above coefficients via fluctuation-dissipation relations. In the following subsections we further discuss  the modification of these coefficients in presence of an external magnetic field.

\subsection{HQ dynamics with finite magnetic field}

Initial arguments in support of the Langevin picture to describe HQ dynamics in the magnetized medium is similar to that of the previous section. In presence of an external magnetic field the heavy quark mass is considered to be sufficiently large, i.e. $M\gg \sqrt{eB}$. The value of the external magnetic field $eB$ will determine the further scale hierarchies, e.g. $M\gg \sqrt{eB} \gg T$ for the Lowest Landau Level dynamics. However, because of the spatial anisotropy introduced by the external magnetic field, we will have a set of two equations for the longitudinal ($z/\sp$) and transverse ($\perp$) momenta
\begin{subequations}
\begin{align}
\frac{dp_z}{dt} &= -\eta_\sp p_z +\xi_z, ~~\langle \xi_z(t)\xi_z(t^\prime)\rangle = \kappa_\sp\delta(t-t^\prime), \\ 
\frac{d\vec p_\perp}{dt} &= -\eta_\perp \vec p_\perp +\vec\xi_\perp, ~~\langle \xi_\perp^i(t)\xi_\perp^j(t^\prime)\rangle = \kappa_\perp\delta_{ij}\delta(t-t^\prime),
\end{align}
\end{subequations}
where $(i,j =x,y) $ and $\vec A_\perp =(A_x,A_y)$ are the transverse components of the momenta, random forces and drag coefficients. The drag and diffusion coefficients are related to each other as: 
\bea \label{eq_eta_kappa}
\eta_\sp &=& \frac{\kappa_\sp}{2MT}, ~~ \eta_\perp = \frac{\kappa_\perp}{2MT}.
\eea
Moreover, similarly as the relativistic case at $B=0$, for the magnetized medium also, within the static limit we can break down $\kappa$ into longitudinal and transverse parts using the rotational symmetry 
\bea
3\kappa = \kappa_\sp + 2\kappa_\perp,
\eea
with 
\begin{subequations}
\begin{align}
&\kappa_\sp = \int d^3q\frac{d~\Gamma(E)}{d^3q}q_\sp^2, \label{msmt_wr_long}\\
&\kappa_\perp = \frac{1}{2}\int d^3q\frac{d~\Gamma(E)}{d^3q}q_\perp^2,
\label{msmt_wr_tran} 
\end{align}
\end{subequations}
where $\frac{d\Gamma(E)}{d^3q}$ can be interpreted as the scattering rate of the HQ via one-gluon exchange with thermal particles per unit volume of momentum transfer $q$. 

On the other hand beyond the static limit we have the finite velocity $\vec{v} = \vec{p}/E$. Now we have to consider the direction of $\vec{v}$ in the context.  

\subsubsection{case 1: $\vec{v} \sp \vec{B}$}

This case is simpler since the magnetic field and the heavy quark point in the same direction, i.e. $z$ direction for our case. So the transport coefficients are given by 
\begin{subequations}
\begin{align}
\frac{1}{2}\frac{d}{dt}\langle (\Delta p_T)^2\rangle \equiv& \kappa_T (p), \\
\frac{d}{dt}\langle (\Delta p_z)^2\rangle \equiv& \kappa_L (p),
\end{align}
\end{subequations}
where $\Delta$ signifies the respective variance of the momentum distributions with the transport coefficients. These   transverse and longitudinal momentum diffusion coefficients are in turn related to scattering rate as follows:  
\begin{subequations}
\label{coeffs_case1}
\begin{align}
\kappa_T (p) =& \frac{1}{2}\int d^3q\frac{d~\Gamma(v)}{d^3q}q_\perp^2, \\
\kappa_L (p) =& \int d^3q\frac{d~\Gamma(v)}{d^3q}q_z^2.
\end{align}
\end{subequations}

\subsubsection{case 2 : $\vec{v} \perp \vec{B}$}

 In this situation as the HQ moves perpendicular to (i.e. $x$ or $y$) the direction of the external anisotropic magnetic field (i.e. $z$), we have three momentum diffusion coefficients (i.e. $\kappa_1, \kappa_2, \kappa_3$) that are different in general: 
\begin{subequations}
\begin{align}
&\frac{d}{dt}\langle (\Delta p_x)^2\rangle \equiv \kappa_1 (p),\\
&\frac{d}{dt}\langle (\Delta p_y)^2\rangle \equiv \kappa_2 (p), \\
&\frac{d}{dt}\langle (\Delta p_z)^2\rangle \equiv \kappa_3 (p),
\end{align}
\end{subequations}
which are explicitly given as
\begin{subequations}
\label{coeffs_case2}
\begin{align}
&\kappa_1 (p) = \int d^3q\frac{d~\Gamma(v)}{d^3q}q_x^2, \\
&\kappa_2 (p) = \int d^3q\frac{d~\Gamma(v)}{d^3q}q_y^2, \\
&\kappa_3 (p) = \int d^3q\frac{d~\Gamma(v)}{d^3q}q_z^2.
\end{align}
\end{subequations}


\section{Computation of the Scattering rate ($\Gamma$)}
\label{sec3}

\begin{figure}
\begin{center}
\includegraphics[scale=0.6]{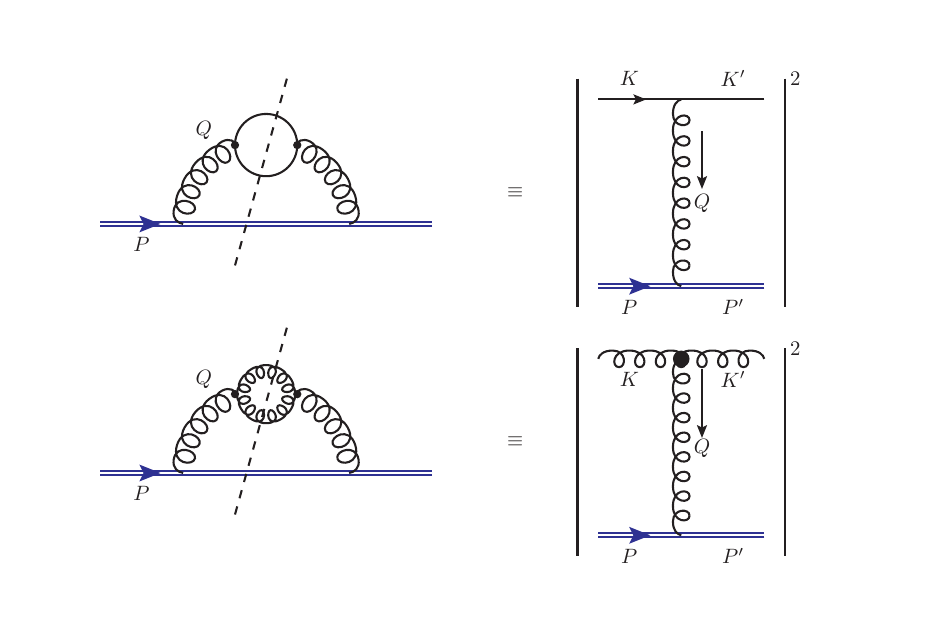}
\caption{The equivalence of the $t$-channel scattering of heavy quarks due to thermally generated light quarks and gluons, $qH\rightarrow qH$ (left) and $gH\rightarrow gH$ (right) are shown, as they can also be expressed as the cut (imaginary) part of the HQ self energy.} 
\label{hq_sqme_alt}
\end{center}
\end{figure}
An effective way of expressing the scattering rate, as proposed by Weldon~\cite{Weldon:1983jn} and demonstrated in Fig. \ref{hq_sqme_alt}, is in terms of the cut/imaginary part of the HQ self energy $\Sigma(P)$, 
\begin{align}
&\Gamma(P\equiv E,{\bf v}) \nn\\
&= -\frac{1}{2E}~\frac{1}{1+e^{-E/T}}~\Tr\left[(\slashed{P}+M)~{\rm Im}\Sigma(p_0+i\epsilon,{\vec{p}})\right].
\label{interaction_rate2}
\end{align}

The advantage of Eq.(\ref{interaction_rate2}) is that one can apply imaginary time formalism of thermal field theory to extract $\Sigma(P)$ including the necessary resummations as we will see soon.

\begin{figure}
\begin{center}
\includegraphics[scale=0.6]{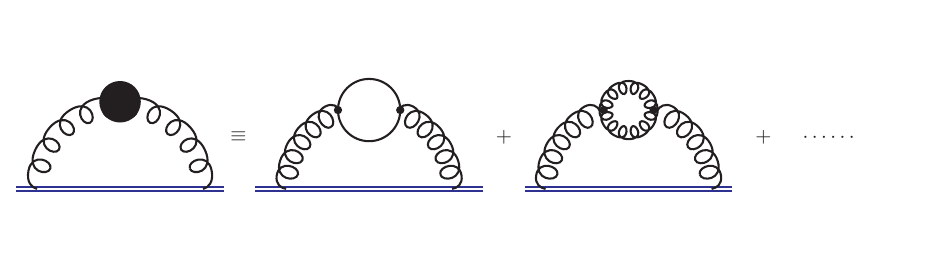}
\caption{Heavy quark self-energy with effective gluon propagator. Resummation takes into account the diagrams for the hard process (same as Fig.\ref{hq_sqme_alt}) among others. } 
\label{hq_htl}
\end{center}
\end{figure}

Now, though the hard contribution of $\Gamma(P)$ comes from cutting the two-loop self energy diagrams shown in Fig. \ref{hq_sqme_alt}. On the other hand, to include the soft contributions, i.e. where the momentum $Q$ flowing through the gluon line is soft, hard thermal loop corrections to the gluon propagator contribute at leading order in $g$. In this case, resummation must be taken into account. So, instead of two separate processes (i.e. $qH\rightarrow qH$ and $gH\rightarrow gH$) depicted in Fig.~\ref{hq_sqme_alt}, we will have an effective gluon propagator which is obtained by summing the geometric series of one-loop self energy corrections proportional to $g^2T^2$ (see Fig.~\ref{hq_htl}). 

\subsection{Scattering rate without magnetic field}

 For $B=0$, one can identify the hard and soft scales as $T$ and $gT$ respectively which enables us to use the HTL approximation assuming $Q\sim gT$. In this case, the effective self-energy for the HQ is given by
\begin{align}
\Sigma(P) &= ig^2 \int \frac{d^4Q}{(2\pi)^4}G^{\mu\nu}(Q) \gamma_\mu\frac{1}{\slashed{P}-\slashed{Q}-M}\gamma_\nu \nn\\
&= -g^2T \sum_{q_0}\int \frac{d^3q}{(2\pi)^3}G^{\mu\nu}(q_0,\vec{q}) \gamma_\mu\frac{1}{\slashed{P}-\slashed{Q}-M}\gamma_\nu,
\end{align}
where $Q\equiv (q_0,\vec{q})$ is the gluonic four-momenta and $G^{\mu\nu}(Q)$ is the HTL gluon propagator in Coulomb gauge, given as 
\bea
G^{\mu\alpha}(Q) = -\frac{\delta^{\mu 0}\delta^{\alpha 0}}{q^2+\Pi_L} + \frac{\delta^{ij}-\hat{q}^i\hat{q}^j}{q^2-q_0^2+\Pi_T}.
\label{t_gp}
\eea
The first term of Eq. (\ref{t_gp}) represents the temporal part of the gluon propagator $G^{00}$ (i.e. it would vanish for $\mu,\alpha \neq 0$) whereas $(i,j)$ in the second term symbolize the spatial components.
$\Pi_L$ and $\Pi_T$ are respectively the longitudinal and transverse coefficients of the HTL gluon self-energies ($\Pi_L$ is also equivalent to the temporal component $\Pi_{00}$ of the HTL gluon self energy $\Pi_{\mn}$), given as
\begin{subequations}
\begin{align}
&\Pi_L = \Pi_{00} = (m_D^2)_g \left\{1-\frac{q_0}{2q}\left[\ln\left(\frac{q+q_0}{q-q_0}\right)-i\pi\right] \right\}, \\
&\Pi_T = (m_D^2)_g \left\{\frac{q_0^2}{2q^2}+\frac{q_0(q^2-q_0^2)}{4q^3}\left[\ln\left(\frac{q+q_0}{q-q_0}\right)-i\pi\right] \right\},
\end{align}
\end{subequations}
with $(m_D)_g$ being the Debye screening mass without magnetic field and $(m_D^2)_g = \frac{g^2N_cT^2}{3}$, $N_c$ being the number of colors.

Now, evaluation of the trace in Eq. (\ref{interaction_rate2}) yields 
\begin{align}
&\Tr\left[(\slashed{P}+M)\Sigma(P)\right] = -4g^2T\sum_{q_0}\int \frac{d^3q}{(2\pi)^3} \frac{1}{(P-Q)^2-M^2}\nn\\
&~~ \Biggl[G_L(Q)\left(p_0^2+p^2-p_0q_0 - \vec{p}\cdot \vec{q}+M^2\right)
+ \nn\\
&~~ 2G_T(Q)\left(p_0^2-p_0q_0+\vec{p}\cdot \vec{q}-( \vec{p}\cdot\hat{q})^2-M^2\right)\Biggr],
\label{tr_zB_ir}
\end{align}
where $G_L$ and $G_T$ are defined as 
\bea
G_L^{-1} &=& q^2 + \Pi_L, \nn\\
 G_T^{-1} &=& q_0^2 -q^2 - \Pi_T. \nn
\eea

To perform the Matsubara sum, the most efficient way is to use the spectral representations~\cite{Pisarski:1987wc} for the fermionic propagators ($P-Q \equiv K$) and the gluonic form factors. Spectral representation of the fermion propagator can be expressed as 
\begin{align}
 &\frac{1}{K^2-M^2} = \frac{1}{k_0^2-E'^2} \nn\\
=& \frac{-1}{2E'}\int\limits_0^{1/T} d\tau'e^{k_0\tau'}\left[n_F(-E')e^{-E'\tau'} -n_F(E')e^{E'\tau'}\right],
\label{spectral_fp}
\end{align}
with $E' = \sqrt{M^2+(\vec{p}-\vec{q})^2}$. Similar procedure for the gluonic form factors yields
\begin{align}
G_{L/T} (Q) = -\int\limits_0^{1/T} \! d\tau e^{q_0\tau} \!\! \int\limits_{-\infty}^{+\infty}\!\! d\omega \rho_{L/T}(\omega,q)[1+n_B(\omega)]e^{-\omega \tau},
\label{spectral_gff}
\end{align}
where $\rho_{L/T}$ are the spectral functions defined as $\rho_{L/T}(\omega,q) = -{\rm Im}G_{L/T}(q_0+i\epsilon,q)/\pi$.

Next, combining Eqs.~(\ref{spectral_fp}) and (\ref{spectral_gff}) in Eq.~(\ref{tr_zB_ir}), evaluating the $\tau,\tau'$ integrals and extracting the imaginary part using the standard formula 
\bea
{\rm Im}\left(\frac{1}{p_0+i\epsilon \mp p}\right) = -i\pi \delta(p_0\mp p),
\eea
one can finally obtain 
\begin{align}
&\Tr\left[(\slashed{P}+M)~{\rm Im}\Sigma(P)\right] \nn\\
=& -4\pi g^2 (1+e^{-p_0/T})\int\frac{d^3q}{(2\pi)^3}\int\limits_{-\infty}^{+\infty}d\omega~[1+n_B(\omega)]~\frac{1}{2E'}\nn\\
&\left\{ [1-n_F(E')]\delta(p_0-E'-\omega)-n_F(E')\delta(p_0+E'-\omega)\right\}\nn\\
&\times \Bigl[ \rho_L(\omega,q) (2p_0^2-p_0\omega-\vec{p}\cdot\vec{q})\nn\\
&~~~~+2\rho_T(\omega,q)(p^2-p_0\omega+\vec{p}\cdot\vec{q}-(\vec{p}\cdot\hat{q})^2)\Bigr].
\end{align}
Next we can simplify the above expression using the assumptions $M,p\gg T$. So, the second $\delta$ function vanishes as $\omega \approx T$. The exponentially suppressed Fermi-Dirac distribution can also be dropped. Using $E' \simeq p_0- \vec{v}\cdot\vec{q}$, the first $\delta$ function becomes $\delta(\omega -\vec{v}\cdot\vec{q})$. Eventually the expression can be written as 
\begin{align}
&\Tr\left[(\slashed{P}+M)~{\rm Im}\Sigma(P)\right] = -4\pi g^2 (1+e^{-p_0/T}) \nn\\
&\int\frac{d^3q}{(2\pi)^3}\int\limits_{-\infty}^{+\infty}d\omega~[1+n_B(\omega)]\frac{1}{2p_0}\delta(\omega -\vec{v}\cdot\vec{q})\nn\\
&\times \Bigl[ \rho_L(\omega,q) (2p_0^2)+2\rho_T(\omega,q)(p^2-(\vec{p}\cdot\hat{q})^2)\Bigr],
\end{align}
which gives the expression for the scattering rate from Eq.(\ref{interaction_rate2}) as~\cite{Braaten:1991jj,Beraudo:2009pe}
\begin{align}
\Gamma(P)&= 2\pi g^2 \!\!\int\!\!\!\frac{d^3q}{(2\pi)^3}\!\!\int\limits_{-\infty}^{+\infty}\!\!d\omega[1+n_B(\omega)]\delta(\omega -\vec{v}\cdot\vec{q})\nn\\
&\left[ \rho_L(\omega,q)+\rho_T(\omega,q)(v^2-(\vec{v}\cdot\hat{q})^2)\right].
\label{sr_zeroB}
\end{align}
This result also reproduces the known result for the damping rate of a static quark~\cite{Pisarski:1993rf} in the static (i.e. $v\rightarrow 0$)  limit. At this point,  we would like to note that even though our HTL approximation within the assumption of $Q\sim gT$ is justified for the calculation of the scattering rate, 
the  $Q\sim T$ scale also becomes relevant for the evaluation of momentum diffusion coefficients ~\cite{Braaten:1991jj}. Hence for the results in  the $eB=0$ case, we have used the same approach as   Ref~\cite{Beraudo:2009pe} where the scattering rate from Eq.~(\ref{sr_zeroB}) has been used to evaluate the momentum diffusion coefficients within the Leading Logarithmic Accuracy (LLA). Within this procedure we need an UV momentum cutoff $q_{max}$ which is to be further discussed in section \ref{sec5}.


\subsection{Scattering rate with finite magnetic field}
Under the presence of a finite magnetic field, the usual counting of scales in Hard Thermal Loop approach gets more complicated due to the new $\sqrt{eB}$ scale. In  the present calculation, 
we consider $T,\sqrt{eB}$ both as hard scales for the loop momenta and $ gT$ as soft scales for the external momenta. More specifically, note that in the effective gluon propagator (shown in Fig.2): for the quark loop there will be the temperature $T$ scale   and additionally the $\sqrt{eB}$ scale will and only will come in via  the Lowest Landau Level for quarks; for  the gluon loop, there will be only the temperature $T$ as the hard scale. We consider the external momentum in gluon propagator to be soft scale $gT$ as usually done in HTL. These scales still respect a hierarchy of $gT \ll T \ll \sqrt{eB}$. 
The effective heavy quark self energy in a magnetized medium is given by, 
\bea
\Sigma(P) = ig^2\int\frac{d^4Q}{(2\pi)^4}\mathcal{D}^{\mn}(Q)\gamma_\mu S_m^s(P-Q)\gamma_\nu.
\eea
In this equation, the fermion propagator in the LLL approximation $S_m^s(P-Q\equiv K)$ is given by~\cite{Schwinger:1951nm,Gusynin:1995nb,Calucci:1993fi}, 
\bea
iS^s_{m}(K)=ie^{-{k_\perp^2}/{|q_fB|}}~~\frac{\slashed{K}_\sp+M}{
K_\sp^2-M^2}(1-i\gamma_1\gamma_2),
\label{prop_sfa}
\eea
where $q_f$ is the fermionic charge for a particular flavor $f$ and $K\equiv (K_\sp,k_\perp)$ is the fermionic four momentum (Details about these $\sp$ and $\perp$ notation can be found in Appendix \ref{appA}). In strong field approximation or in LLL, $eB \gg k_\perp^2$, an effective dimensional reduction from $(3+1)$ to $(1+1)$ takes place~\cite{Gusynin:1995nb,Calucci:1993fi}. 
We note that the LLL approximation works best under the condition $\frac{eB}{M} \gg T$. 

It shall be noted that  there have been considerable new developments in the exploration of the thermo-magnetic corrections to the correlation functions. Recently the thermo-magnetic correction to the quark-gluon vertex has  been computed in the weak magnetic field limit within the HTL approximation~\cite{Ayala:2014uua,Haque:2017nxq}. Also there are several recent studies on the general structures of the fermion and gauge boson self-energies with propagators at finite temperature and in presence of an external magnetic field~\cite{Shabad:2010hx,Hattori:2012je,Bordag:2008wp,Chao:2014wla,Mueller:2014tea,Das:2017vfh,Ayala:2018ina,Karmakar:2018aig,Ayala:2020wzl,Ayala:2021lor}. These studies vary in their approach by their choice of the independent tensor structures for constructing  the two-point correlation functions. Out of these choices we have chosen the effective gluon propagator in a hot and magnetized medium from~\cite{Karmakar:2018aig}, i.e.,
\begin{align}
&\mathcal{D}^{\mn}(Q)=\frac{\xi Q^{\mu}Q^{\nu}}{Q^4}+\frac{(Q^2-d_3) \Delta_1^{\mn}}{(Q^2-d_1)(Q^2-d_3)-d_4^2}+\frac{\Delta_2^{\mn}}{Q^2-d_2}\nn\\
&+\frac{(Q^2-d_1) \Delta_3^{\mn}}{(Q^2-d_1)(Q^2-d_3)-d_4^2} +\frac{d_4 \Delta_4^{\mn}}{(Q^2-d_1)(Q^2-d_3)-d_4^2},
\label{gauge_prop}
\end{align}
with
\begin{subequations}
\begin{align}
d_1(Q) &= \Delta_1^{\mn}\Pi_{\mn}(Q), \label{ff_d1} \\
d_2(Q) &= \Delta_2^{\mn}\Pi_{\mn}(Q), \label{ff_d2} \\
d_3(Q) &= \Delta_3^{\mn}\Pi_{\mn}(Q), \label{ff_d3} \\
d_4(Q) &= \frac{1}{2}\Delta_4^{\mn}\Pi_{\mn}(Q) \label{ff_d4},
\end{align}
\end{subequations} 
and 
\begin{subequations}
\begin{align}
\Delta_1^{\mu\nu} &= \frac{1}{\bar{u}^2} \bar{u}^\mu\bar{u}^\nu, \label{D1munu}\\
\Delta_2^{\mu\nu}  &=g_{\perp}^{\mn}-\frac{Q^{\mu}_{\perp}Q^{\nu}_{\perp}}{Q_{\perp}^2}, \label{D2munu}\\
\Delta_3^{\mu\nu}  &=  \frac{{\bar n}^\mu {\bar n}^\nu}{\bar n^2}, \label{D3munu}\\
\Delta_4^{\mu\nu} &= \frac{\bar u^{\mu}\bar n^{\nu}+\bar u^{\nu}\bar n^{\mu}}{\sqrt{\bar u^2}\sqrt{\bar n^2}},
\end{align}
\end{subequations}
where $u^\mu$ is the heat bath velocity and $n^\mu$ is defined uniquely as the projection of the electromagnetic field tensor $F^{\mn}$ along $u^\mu$. Details about the construction of the tensor structure and the notations of $\bar{u}^\mu, \bar{n}^\nu, g_\perp^{\mn}$ etc. are given in Appendix \ref{appA}.  $\Pi_{\mn}(Q)$ is the HTL gluon self energy in a strongly magnetized hot medium which is a combination of the Yang-Mills contribution $\Pi^g_{\mn}$ and fermionic loop contribution $\Pi^s_{\mn}$ within LLL approximation. The expressions for $\Pi^s_{\mn}$, $\Pi^g_{\mn}$ and the evaluation of $d_i(Q)$'s within the LLL approximation are given in Appendix \ref{appB}. 

Next we evaluate the trace required for the scattering rate, i.e. 
\begin{align}
&\Tr\left[(\slashed{P}+M)~\Sigma(P)\right] = ig^2\int\frac{d^4Q}{(2\pi)^4} \frac{e^{-{k_\perp^2}/{|q_fB|}}}{K_\sp^2-M^2} \nn\\
&\times\sum\limits_{i=1}^4\mathcal{J}_i~\Tr\left[(\slashed{P}+M)\Delta_i^{\mn}\gamma_\mu (\slashed{K}_\sp+M)(1-i\gamma_1\gamma_2)\gamma_\nu\right], 
\end{align}
where we are working in a gauge with vanishing gauge parameters. The coefficients $\mathcal{J}_i$'s are given as, 
\begin{subequations}
\begin{align}
\mathcal{J}_1 &= \frac{(Q^2-d_3)}{(Q^2-d_1)(Q^2-d_3)-d_4^2}, \\
\mathcal{J}_2 &= \frac{1}{(Q^2-d_2)}, \\
\mathcal{J}_3 &=\frac{(Q^2-d_1)}{(Q^2-d_1)(Q^2-d_3)-d_4^2}, \\
\mathcal{J}_4 &= \frac{d_4}{(Q^2-d_1)(Q^2-d_3)-d_4^2}.
\end{align}
\label{mathcaljs}
\end{subequations} 

We can now evaluate the individual traces as 
\begin{subequations}
\begin{align}
&\Tr\left[(\slashed{P}+M)\Delta_1^{\mn}\gamma_\mu(\slashed{K}_\sp+M)(1-i\gamma_1\gamma_2)\gamma_\nu\right]\nn\\
 =& \frac{4}{\bar{u}^2} \left[2(\bar{u}\cdot K)_\sp (\bar{u}\cdot P)-\bar{u}^2\left((K\cdot P)_\sp-M^2\right)\right]\nn\\
 =& \frac{4}{\bar{u}^2} \Bigl[2\left(p_0 - q_0 \left(1+\frac{(P\cdot Q)_\sp-Q_\sp^2}{Q^2}\right)\right) \nn\\
&\times\left(p_0-q_0\frac{P\cdot Q}{Q^2}\right)+\bar{u}^2\left(M^2 - P_\sp^2 - p_3q_3 + p_0q_0\right)\Bigr]\nn\\
=&A_1 + q_0 B_1,
\end{align}
where 
\begin{equation}
A_1 = \frac{4}{\bar{u}^2} \Bigl[2p_0^2+\bar{u}^2\left(M^2 - P_\sp^2 - p_3q_3\right)\Bigr]
\label{A1_final}
\end{equation}
and $B_1$ represents rest of the $q_0$ dependent terms.

\begin{align}
&\Tr\left[(\slashed{P}+M)\Delta_2^{\mn}\gamma_\mu(\slashed{K}_\sp+M)(1-i\gamma_1\gamma_2)\gamma_\nu\right]\nn\\
&~~= -4 (K\cdot P)_\sp + 4M^2\nn\\
&~~= 4\left(M^2 - P_\sp^2 - p_3q_3 + p_0q_0\right)\nn\\
&~~= A_2 + q_0 B_2,  
\end{align}
with 
\begin{equation}
A_2 = 4\left(M^2 - P_\sp^2 - p_3q_3\right)
\label{A2_final}
\end{equation}
and $B_2$ represents rest of the $q_0$ dependent term.

\begin{align}
&\Tr\left[(\slashed{P}+M)\Delta_3^{\mn}\gamma_\mu(\slashed{K}_\sp+M)(1-i\gamma_1\gamma_2)\gamma_\nu\right]\nn\\
 =& \frac{4}{\bar{n}^2} \left[2(\bar{n}\cdot K)_\sp (\bar{n}\cdot P)-\bar{n}^2\left((K\cdot P)_\sp-M^2\right)\right]\nn\\
 =& \frac{4}{\bar{n}^2} \Bigl[2\left(-k_3+\frac{q_0q_3k_0}{q^2}-\frac{q_3}{q^2}((P\cdot Q)_\sp-Q_\sp^2)\right)\nn\\
& \times\left(-p_3+\frac{q_0q_3p_0}{q^2}-\frac{q_3}{q^2}(P\cdot Q)\right)\nn\\
&~+\bar{n}^2\left(M^2 - P_\sp^2 - p_3q_3 + p_0q_0\right)\Bigr]\nn\\
 =&A_3 + q_0 B_3,
\end{align}
with 
\begin{align}
A_3 &= 4\Bigl[\frac{2k_3q_3}{q^2}(\vec{p}\cdot\vec{q})+M^2 - p_0^2- p_3k_3\Bigr]
\label{A3_final}
\end{align}
and $B_3$ represents rest of the $q_0$ dependent terms.

\begin{align}
&\Tr\left[(\slashed{P}+M)\Delta_4^{\mn}\gamma_\mu(\slashed{K}_\sp+M)(1-i\gamma_1\gamma_2)\gamma_\nu\right]\nn\\
&= \frac{4}{\sqrt{\bar{u}^2}\sqrt{\bar{n}^2}} \big[(\bar{u}\cdot K)_\sp (\bar{n}\cdot P)+(\bar{n}\cdot K)_\sp (\bar{u}\cdot P)\nn\\
&-2(\bar{n}\cdot\bar{u})\left((K\cdot P)_\sp-M^2\right)\big] \nn\\
&= \frac{4}{\sqrt{\bar{u}^2}\sqrt{\bar{n}^2}} \Bigl[\left(p_0 - q_0 \left(1+\frac{(P\cdot Q)_\sp-Q_\sp^2}{Q^2}\right)\right) \times \nn\\
&\left(-p_3+\frac{q_0q_3p_0}{q^2}-\frac{q_3}{q^2}(P\cdot Q)\right)\nn\\ 
&+\left(-k_3+\frac{q_0q_3k_0}{q^2}-\frac{q_3}{q^2}((P\cdot Q)_\sp-Q_\sp^2)\right)\nn\\
&\left(p_0-q_0\frac{P\cdot Q}{Q^2}\right)\Bigr]\nn\\
 &= A_4 + q_0 B_4,
\end{align}
with 
\begin{align}
A_4 & = \frac{4p_0}{\sqrt{\bar{u}^2}\sqrt{\bar{n}^2}} \big[\left(-p_3+\frac{q_3}{q^2}(\vec{p}\cdot\vec{q})\right)+k_3\bar{n}^2\big]
\label{A4_final}
\end{align}
and $B_4$ represents rest of the $q_0$ dependent terms. 
\end{subequations} 

Next we compute the sum over $q_0$, for which we introduce the spectral representations for the propagators. The spectral representation for the fermionic part can be obtained using 
\begin{align}
&\frac{1}{K_\sp^2-M^2} =-\frac{1}{2E'_\sp}\times\nn\\
& \int\limits_0^{1/T}d\tau' e^{k_0\tau'} \left[(1-n_F(E'_\sp))e^{-E'_\sp\tau'} - n_F(E'_\sp)e^{E'_\sp\tau'} \right],
\end{align}
with $E'_\sp = \sqrt{k_3^2+M^2}$. On the other hand, pieces from the effective gluon propagator appearing in Eqs. (\ref{mathcaljs}) can be represented as
\begin{align}
&\mathcal{J}_i = - \int\limits_0^{1/T}d\tau~ e^{q_0\tau} \int\limits_{-\infty}^{+\infty}~ d\omega~ \rho_i(\omega,q)~\left[1+n_B(\omega)\right]~e^{-\omega\tau}.
\end{align}
The corresponding spectral functions are given by
\begin{align}
&\rho_i(\omega,q) = -\frac{1}{\pi} {\rm Im}\left(\mathcal{J}_i\Big |_{q_0 = \omega +i\epsilon}\right).
\end{align}
Detailed evaluations of these spectral functions are given in Appendix \ref{appC}. Now the sum over $q_0$ can be evaluated from the combination of the integrals over $\tau$ and $\tau'$, using 
\begin{subequations}
\begin{align}
\sum_{q_0} e^{q_0(\tau-\tau')} =& \delta (\tau-\tau'), \\
\sum_{q_0} q_0 ~e^{q_0(\tau-\tau')} =& \delta' (\tau-\tau').
\end{align}
\end{subequations}
This subsequently yields
\begin{widetext}
\begin{align}
\Tr\left[(\slashed{P}+M)~\Sigma(P)\right] =&~ ig^2\int\frac{d^4Q}{(2\pi)^4}\frac{e^{-{k_\perp^2}/{|q_fB|}}}{K_\sp^2-M^2}  \sum\limits_{i=1}^4 \mathcal{J}_i~\left[A_i + q_0 B_i\right] \nn\\
=& -g^2T\sum\limits_{i=1}^4\int\frac{d^3q}{(2\pi)^3} e^{-{k_\perp^2}/{|q_fB|}}\int\limits_{-\infty}^{+\infty} d\omega\left[1+n_B(\omega)\right]\int\limits_0^{1/T}d\tau'\int\limits_0^{1/T}d\tau ~e^{p_0\tau'-\omega\tau} \nn\\
&\times\sum_{q_0}e^{q_0(\tau-\tau')}\left[A_i + q_0 B_i\right]~~\frac{\rho_i(\omega,q)}{2E'_\sp}\left[(1-n_F(E'_\sp))e^{-E'_\sp\tau'} - n_F(E'_\sp)e^{E'_\sp\tau'}\right] \nn\\
=& -g^2T\sum\limits_{i=1}^4\int\frac{d^3q}{(2\pi)^3} e^{-{k_\perp^2}/{|q_fB|}}\int\limits_{-\infty}^{+\infty} d\omega\left[1+n_B(\omega)\right]\frac{\rho_i(\omega,q)}{2E'_\sp}\left(A_iP_1 + B_iP_2 \right),
\end{align}
\end{widetext}
where expressions for $P_1$ and $P_2$ are given below. 
\begin{align}
P_1 =& \int\limits_0^{1/T}d\tau'\int\limits_0^{1/T}d\tau ~e^{p_0\tau'-\omega\tau} \delta (\tau-\tau')\nn\\
&~\times\left[(1-n_F(E'_\sp))e^{-E'_\sp\tau'} - n_F(E'_\sp)e^{E'_\sp\tau'}\right]\nn\\
 =& \int\limits_0^{1/T}d\tau ~e^{(p_0-\omega)\tau}\left[(1-n_F(E'_\sp))e^{-E'_\sp\tau} - n_F(E'_\sp)e^{E'_\sp\tau}\right] \nn\\
=&-\sum_{\sigma=\pm 1} \frac{\sigma~n_F(\sigma E'_\sp)}{p_0+\sigma E'_\sp-\omega}\left(e^{(p_0+\sigma E'_\sp-\omega)/T}-1\right).
\label{P1_final}
\end{align}
Similarly for $P_2$ we obtain 
\begin{align}
P_2 =& \int\limits_0^{1/T}d\tau'\int\limits_0^{1/T}d\tau ~e^{p_0\tau'-\omega\tau} \delta'(\tau-\tau')\nn\\
&\left[(1-n_F(E'_\sp))e^{-E'_\sp\tau'} - n_F(E'_\sp)e^{E'_\sp\tau'}\right]\nn\\
=& -\!\!\!\!\!\int\limits_0^{1/T}d\tau ~\frac{d}{d\tau}~e^{(p_0-\omega)\tau}\left[(1-n_F(E'_\sp))e^{-E'_\sp\tau} - n_F(E'_\sp)e^{E'_\sp\tau}\right] \nn\\
=& \sum_{\sigma=\pm 1} \sigma~n_F(\sigma E'_\sp)\left(e^{(p_0+\sigma E'_\sp-\omega)/T}-1\right).
\label{P2_final}
\end{align}

At the discrete imaginary energies $p_0 = i(2n+1)\pi T$, we can eliminate the $p_0$ from the exponent as $e^{p_0/T} = -1$. Then after analytic continuation from $p_0\rightarrow E + i\epsilon$, the imaginary part of $\Sigma$ comes from the energy denominator as 
\begin{align}
{\rm Im}~\left(\frac{1}{p_0+\sigma E'_\sp - \omega}\right)\Big |_{p_0\rightarrow E+i\epsilon}= -i\pi~\delta(E + \sigma E'_\sp - \omega).
\end{align}
As Eq.~(\ref{P2_final}) implies, $P_2$ doesn't correspond to any imaginary parts. Collecting all these finally we can write down the evaluation for the trace as  
\begin{align}
&\Tr\left[(\slashed{P}+M)~{\rm Im}\Sigma(p_0+i\epsilon, \vec{p})\right] \nn \\
=&~ \pi g^2T\sum\limits_{i=1}^4 \int\frac{d^3q}{(2\pi)^3} e^{-{k_\perp^2}/{|q_fB|}}\nn\\
&\times \int\limits_{-\infty}^{+\infty} d\omega\left[1+n_B(\omega)\right]\frac{\rho_i(\omega,q)A_i}{2E'_\sp} \nn\\
&\times\sum_{\sigma=\pm 1} \sigma~n_F(\sigma E'_\sp)\left(e^{(\sigma E'_\sp-\omega)/T}+1\right)\delta(E + \sigma E'_\sp - \omega) \nn\\
=&~ \pi g^2T\left(e^{-E/T}+1\right)\sum\limits_{i=1}^4 \int\frac{d^3q}{(2\pi)^3} e^{-{k_\perp^2}/{|q_fB|}}\nn\\
&\times \int\limits_{-\infty}^{+\infty} d\omega\left[1+n_B(\omega)\right]\frac{\rho_i(\omega,q)A_i}{2E'_\sp}\nn\\
&\times\sum_{\sigma=\pm 1} \sigma~n_F(\sigma E'_\sp)~\delta(E + \sigma E'_\sp - \omega).
\end{align}

Eventually using Eq.~(\ref{interaction_rate2}), we can obtain the final expression for the interaction rate $\Gamma(E,\vec{v})$ for a particular flavor $f$ as 
\begin{align}
\Gamma(E,\vec{v}) =& -\frac{\pi g^2T}{2E} \sum\limits_{i=1}^4 \int\frac{d^3q}{(2\pi)^3} e^{-{k_\perp^2}/{|q_fB|}}\nn\\
&\times \int\limits_{-\infty}^{+\infty} d\omega\left[1+n_B(\omega)\right]\frac{\rho_i(\omega,q)A_i}{2E'_\sp}\nn\\
&\times\sum_{\sigma=\pm 1} \sigma~n_F(\sigma E'_\sp)~\delta(E + \sigma E'_\sp - \omega).
\end{align}

We can now simplify the expression for the interaction rate a bit further using the scale hierarchy $M\gg \sqrt{eB} \gg T$. As $E \sim E'_\sp \sim M$, so the delta function $\delta(E + E'_\sp-\omega)$ cannot contribute for $\omega \leq T$. Also, the Fermi-Dirac disctribution $n_F(E'_\sp)$ will be exponentially suppressed. These changes subsequently simplify the expression of the scattering rate as 
\begin{align}
&\Gamma(E,\vec{v}) = \frac{\pi g^2T}{2E} \sum\limits_{i=1}^4 \int\frac{d^3q}{(2\pi)^3} e^{-{k_\perp^2}/{|q_fB|}}\nn\\
&\int\limits_{-\infty}^{+\infty} d\omega\left[1+n_B(\omega)\right]\frac{\rho_i(\omega,q)A_i}{2E'_\sp} \delta(E - E'_\sp - \omega).
\end{align}

\section{Energy loss and momentum diffusion coefficients for heavy quark in a strongly magnetized medium}
\label{sec4}

\subsection{case 1 : $\vec{v} \sp \vec{B}$}
\label{case1_exprs}

For this case we only have a nonzero $p_3 (p_z)$ whereas $p_1 (p_x)=p_2 (p_y)=0$. Hence $E=\sqrt{p_3^2+M^2}$ and one can express $E'_\sp = \sqrt{(p_3-q_3)^2+M^2}$ in terms of $E$ by expanding 
\begin{equation}
E'_\sp \approx E - \frac{p_3q_3}{E} = E - v_3q_3
\end{equation}
which results in 
\begin{align}
&\Gamma(E,v_3) = \frac{\pi g^2T}{4E} \sum\limits_{i=1}^4 \int\frac{d^3q}{(2\pi)^3} e^{-{q_\perp^2}/{|q_fB|}}\nn\\
&\int\limits_{-\infty}^{+\infty} d\omega\left[1+n_B(\omega)\right]\frac{\rho_i(\omega,q)A_i^{(1)}}{(E-v_3q_3)} \delta(\omega-v_3q_3),
\end{align}
where $A_i^{(1)}$ corresponds to $A_i$'s from Eqs.~(\ref{A1_final}), (\ref{A2_final}), (\ref{A3_final}) and (\ref{A4_final}) with $p_1=p_2 =0$.

Next within this case we can write down the expressions for the energy loss and the respective momentum diffusion coefficients using Eq.~(\ref{coeffs_case1}). The energy loss will be given as 
\begin{align}
&\frac{dE}{dx} = \frac{\pi g^2T}{4Ev_3} \sum\limits_{i=1}^4 \int\frac{d^3q}{(2\pi)^3} e^{-{q_\perp^2}/{|q_fB|}}\nn\\
&\int\limits_{-\infty}^{+\infty} d\omega\left[1+n_B(\omega)\right]\omega\frac{\rho_i(\omega,q)A_i^{(1)}}{(E-v_3q_3)} \delta(\omega-v_3q_3),
\end{align}
Now, as the spectral functions are odd functions, we can replace the factor $(1+n_B(\omega))$ with its even part, as 
\begin{equation}
(1+n_B(\omega)) \rightarrow \frac{(1+n_B(\omega))+(1+n_B(-\omega))}{2}=\frac{1}{2}\nn
\end{equation}
resulting 
\begin{align}
\frac{dE}{dx} =& \frac{\pi g^2T}{8Ev_3} \sum\limits_{i=1}^4 \int\frac{d^3q}{(2\pi)^3} e^{-{q_\perp^2}/{|q_fB|}}\nn\\
&\int\limits_{-\infty}^{+\infty} d\omega~\omega~\frac{\rho_i(\omega,q)A_i^{(1)}}{(E-v_3q_3)} \delta(\omega-v_3q_3).
\end{align}

Similarly the transverse momentum diffusion coefficient will be given as 
\begin{align}
&\kappa_T(p_3) = \frac{\pi g^2T}{8E} \sum\limits_{i=1}^4 \int\frac{d^3q}{(2\pi)^3} q_\perp^2 e^{-{q_\perp^2}/{|q_fB|}}\nn\\
&\int\limits_{-\infty}^{+\infty} d\omega\left[1+n_B(\omega)\right]\frac{\rho_i(\omega,q)A_i^{(1)}}{(E-v_3q_3)} \delta(\omega-v_3q_3).
\end{align}
Again as the spectral function is odd, we choose to replace the factor $(1+n_B(\omega))$ with its odd part, as 
\begin{equation}
(1+n_B(\omega)) \rightarrow \frac{(1+n_B(\omega))-(1+n_B(-\omega))}{2}=\frac{1}{2}\coth\frac{\omega}{2T}\nn
\end{equation}
resulting 
\begin{align}
&\kappa_T(p_3) = \frac{\pi g^2T}{16E} \sum\limits_{i=1}^4 \int\frac{d^3q}{(2\pi)^3} q_\perp^2~ e^{-{q_\perp^2}/{|q_fB|}}\nn\\
&\int\limits_{-\infty}^{+\infty} d\omega\coth\left(\frac{\omega}{2T}\right)\frac{\rho_i(\omega,q)A_i^{(1)}}{(E-v_3q_3)} \delta(\omega-v_3q_3).
\label{kappaT_case1_final}
\end{align}

Finally the longitudinal momentum diffusion coefficient will be given as 
\begin{align}
&\kappa_L(p_3) = \frac{\pi g^2T}{8E} \sum\limits_{i=1}^4 \int\frac{d^3q}{(2\pi)^3} q_3^2 ~e^{-{q_\perp^2}/{|q_fB|}}\nn\\
&\int\limits_{-\infty}^{+\infty} d\omega\coth\left(\frac{\omega}{2T}\right)\frac{\rho_i(\omega,q)A_i^{(1)}}{(E-v_3q_3)} \delta(\omega-v_3q_3).
\end{align}

One may take the $v_3 \to 0$ limit to obtain results for the case of a static heavy quark. It may be noted that the static limit results here differ from that obtained in \cite{Fukushima:2015wck}. The origin of such difference comes from the different treatment of the gluon self energy, for which we include both quark and gluon loop contributions while \cite{Fukushima:2015wck} considers only the quark loop. In appendix \ref{appD} we have shown that excluding the gluon loop contribution our results agree with that of \cite{Fukushima:2015wck}.

\subsection{case 2 : $\vec{v} \perp \vec{B}$}

For this case we have nonzero $p_1$ and/or $p_2$ whereas $p_3=0$. Hence $E=\sqrt{p_\perp^2 + M^2}$ and $E'_\sp = \sqrt{q_3^2 + M^2}$. Following similar steps as in subsection \ref{case1_exprs} and using Eq.~(\ref{coeffs_case2}), we can straightway write down the expressions for the energy loss and the diffusion momentum coefficients as

\begin{align}
\frac{dE}{dx} =& \frac{\pi g^2T}{8Ev} \sum\limits_{i=1}^4 \int\frac{d^3q}{(2\pi)^3} e^{-{k_\perp^2}/{|q_fB|}}\nn\\
&\int\limits_{-\infty}^{+\infty} d\omega~\omega~\frac{\rho_i(\omega,q)A_i^{(2)}}{E'_\sp} \delta(\omega-E+E'_\sp), \\
\kappa_1(p) =& \frac{\pi g^2T}{8E} \sum\limits_{i=1}^4 \int\frac{d^3q}{(2\pi)^3} q_1^2~ e^{-{k_\perp^2}/{|q_fB|}}\nn\\
&\int\limits_{-\infty}^{+\infty} d\omega\coth\left(\frac{\omega}{2T}\right)\frac{\rho_i(\omega,q)A_i^{(2)}}{E'_\sp} \delta(\omega-E+E'_\sp), \\
\kappa_2(p) =& \frac{\pi g^2T}{8E} \sum\limits_{i=1}^4 \int\frac{d^3q}{(2\pi)^3} q_2^2~ e^{-{k_\perp^2}/{|q_fB|}}\nn\\
&\int\limits_{-\infty}^{+\infty} d\omega\coth\left(\frac{\omega}{2T}\right)\frac{\rho_i(\omega,q)A_i^{(2)}}{E'_\sp} \delta(\omega-E+E'_\sp),\\
\kappa_3(p) =& \frac{\pi g^2T}{8E} \sum\limits_{i=1}^4 \int\frac{d^3q}{(2\pi)^3} q_3^2 ~e^{-{k_\perp^2}/{|q_fB|}}\nn\\
&\int\limits_{-\infty}^{+\infty} d\omega\coth\left(\frac{\omega}{2T}\right)\frac{\rho_i(\omega,q)A_i^{(2)}}{E'_\sp} \delta(\omega-E+E'_\sp).
\end{align}

Here $A_i^{(2)}$ corresponds to $A_i$'s from Eqs.~(\ref{A1_final}), (\ref{A2_final}), (\ref{A3_final}) and (\ref{A4_final}) with $p_3 =0$.

\section{Results}
\label{sec5}

\begin{figure*}
\begin{center}
\includegraphics[scale=0.5]{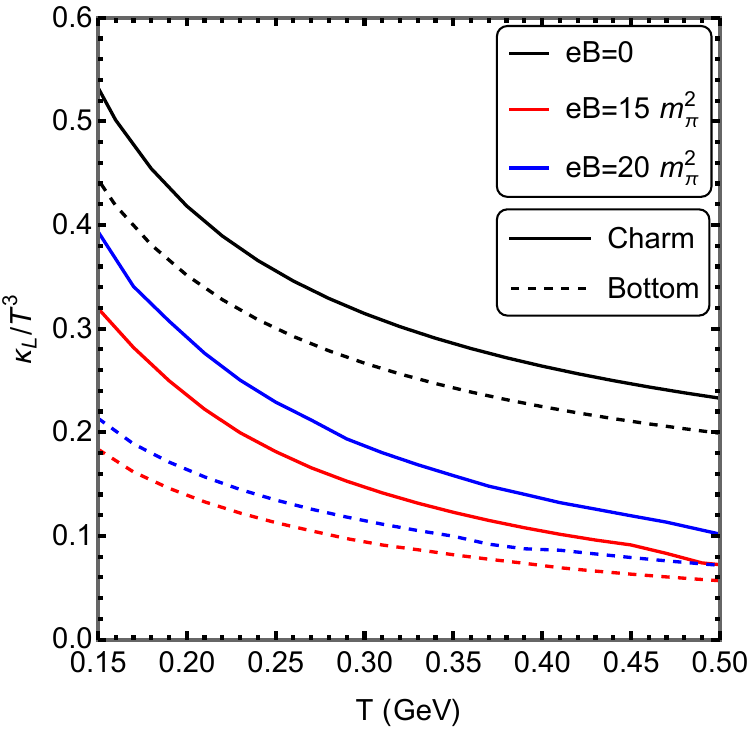}
\hspace{1cm}
\includegraphics[scale=0.5]{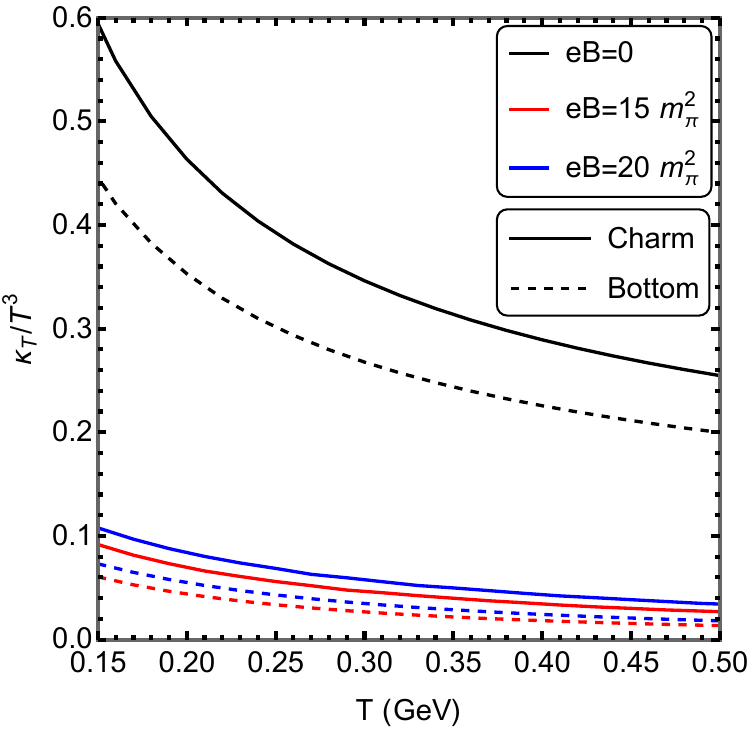}
\caption{Variation of the scaled charm (solid lines) and bottom (dashed lines) quark momentum diffusion coefficients (for $\vec{v} \sp \vec{B}$) with temperature for three different values of external magnetic field, i.e. $eB=0, 15m_\pi^2, 20m_\pi^2$. Left panel shows the variation of the scaled longitudinal components $\kappa_L$, whereas right panel shows the same for the scaled transverse components $\kappa_T$. Charm and bottom quark masses $M$ are specified in the text and HQ momentum $p$ is taken to be 1 GeV. } 
\label{kappavT_LT_scaled}
\end{center}
\end{figure*}

\begin{figure*}[!hbt]
\begin{center}
\includegraphics[scale=0.5]{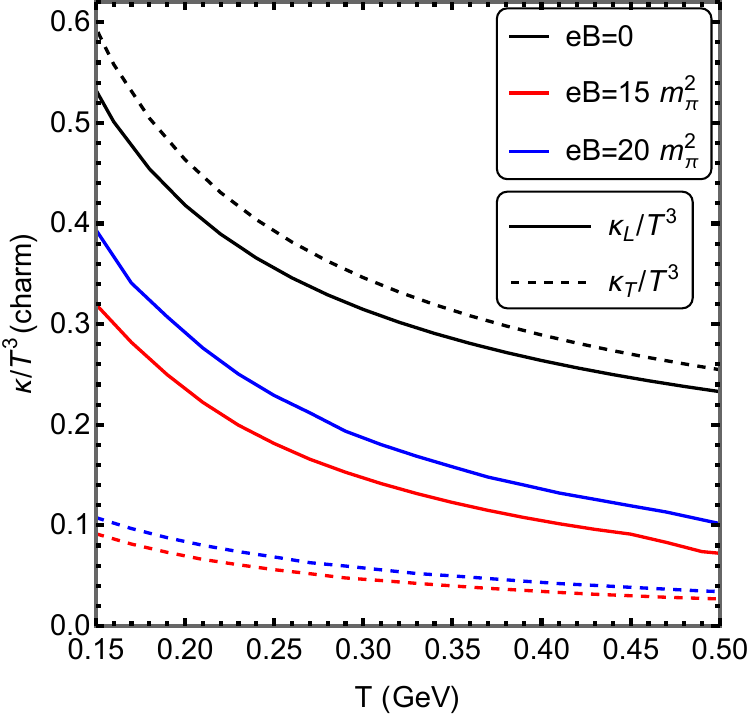}\hspace{1cm}\includegraphics[scale=0.5]{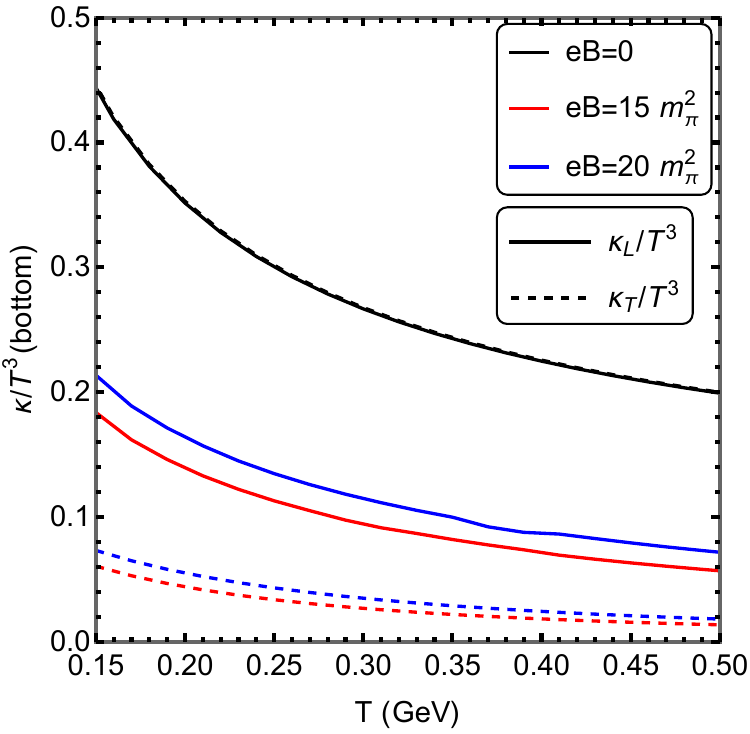}
\caption{Variation of the scaled HQ longitudinal (solid lines) and transverse (dashed lines) momentum diffusion coefficients (for $\vec{v} \sp \vec{B}$) with temperature for three different values of external magnetic field, i.e. $eB=0, 15m_\pi^2, 20m_\pi^2$ and for both charm (left panel) and bottom (right panel) quarks. Charm and bottom quark masses $M$ are specified in the text and HQ momentum $p$ is taken to be 1 GeV.  } 
\label{kappavT_cb_scaled}
\end{center}
\end{figure*}

\begin{figure*}
\begin{center}
\includegraphics[scale=0.5]{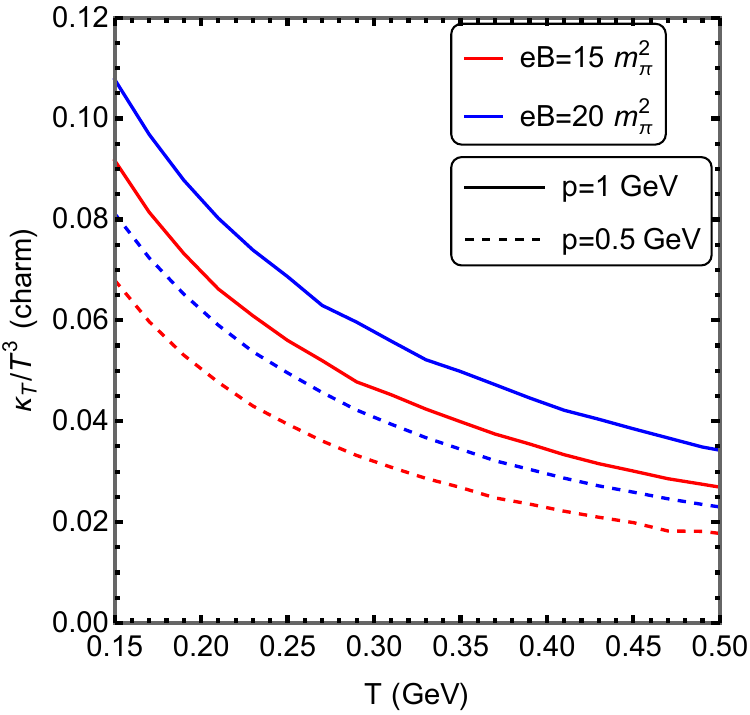}
\caption{Variation of the scaled HQ transverse momentum diffusion coefficient (for $\vec{v} \sp \vec{B}$) with temperature for two different values of external magnetic field and two different values of the HQ momentum $p$. Heavy quark masses $M$ are specified in the text.  } 
\label{kappaT_scaled_vp}
\end{center}
\end{figure*}

In the following subsections we discuss our findings for different momentum diffusion coefficients for heavy charm and bottom quarks moving through a strongly magnetized hot medium. For the numerical calculations, we have used the self-consistent one-loop running coupling $g(T)$, given as 
\begin{equation}
g(\Lambda) = \left[\frac{48\pi^2}{(33-2N_f)\ln\left( \frac{\Lambda^2}{\Lambda_{\overline{MS}}^2}\right)}\right]^{1/2},
\end{equation}
where $\Lambda$ and $\Lambda_{\overline{MS}}$ are the renormalization and the $\overline{MS}$ scales.  The parameter $\Lambda_{\overline{MS}}$ needs to be fixed from a reference point and we follow the lattice calculation in  Ref.~\cite{Bazavov:2012ka} giving the value of $\alpha_s =g^2/4\pi = 0.326$ for the renormalization scale $\Lambda=1.5$ GeV, which thus suggests a value of 
$\Lambda_{\overline{MS}} = 176$ MeV. Given this parameter, we can then obtain the coupling constant at any temperature $T$ by identifying $\Lambda \to 2\pi T$ in the above running coupling formula. We note in passing that  there are recent advances in the determination of $\alpha_s$ while taking into account the magnetic effects~\cite{Ayala:2014uua,Ayala:2016bbi,Ayala:2018wux,Ayala:2019nna}, which may be interesting to incorporate in a future study. 

\subsection{case 1 : $\vec{v} \sp \vec{B}$}
\label{case1_results}

For the $\vec{v} \sp \vec{B}$ case we have only one anisotropic direction which gives rise to two different momentum coefficients, namely $\kappa_L$ and $\kappa_T$, representating the longitudinal and transverse components. In this case the heavy quark momentum is only nonvanishing in the $\vec{B}$ direction, which we have chosen to be $z$. In the following we discuss our results for $\kappa_L$ and $\kappa_T$ for charm and bottom quarks (mass $M=1.28$ GeV and  $M=4.18$ GeV respectively) moving parallel to an external magnetic field along the $z$ direction. For most of our numerical results, we have chosen the HQ momentum $p$ to be 1 GeV. Such a choice allows us to clearly go beyond the static limit while still  maintaining the scale hierarchy of $T \ll p\lesssim M  $ in consistency with our derivations. While studying the HQ momentum dependence of the momentum diffusion coefficients, we also show results for a lower value of $p$, i.e. $0.5$ GeV in comparison with that of $1$ GeV.  We will discuss more about this later in this section. We have also compared our finite $eB$ results with the $eB=0$ results obtained from Ref.~\cite{Beraudo:2009pe}. We have chosen the Ultra-Violate (UV) cut-off $q_{max}$ required for the $eB=0$ case as $q_{max} = 3.1T g(T)^{1/3}$, as discussed in Ref.~\cite{Beraudo:2009pe}. We would also like to note at this point that for finite $eB$ calculations, an UV cut-off like $q_{max}$ is not necessary due to the $e^{-k_\perp^2/|q_fB|}$ factor appearing from the fermion propagator in a magnetized medium. 

In Fig. \ref{kappavT_LT_scaled} we have plotted the variations of scaled longitudinal and transverse momentum coefficients, $\kappa_L/T^3$ (left panel) and $\kappa_T/T^3$ (right panel) with temperature. In both the plots we have shown the variations of both charm (solid lines) and bottom (dashed lines) quarks for three different values of magnetic field, i.e. $eB=0, 15m_\pi^2$ and $20 m_\pi^2$. It can be observed from Fig. \ref{kappavT_LT_scaled} that for increasing magnetic field, both longitudinal and transverse components of the momentum diffusion coefficients have increased. Although when compared with the $eB=0$ case, the values for $\kappa_T$ appear to be significantly reduced by finite magnetic fields.

Fig. \ref{kappavT_cb_scaled} shows a similar variation as in Fig. \ref{kappavT_LT_scaled}, but this time we show two different plots for charm (left panel) and bottom (right panel) quarks and in each plots we present both $\kappa_L$ (solid lines) and $\kappa_T$ (dashed lines) together. As was also evident from Fig. \ref{kappavT_LT_scaled}, interestingly we observe that though for finite $eB$, values of $\kappa_L$ are significantly higher than $\kappa_T$, for $eB=0$ the situation is different. For charm quark (left panel) values of $\kappa_T$ at $eB=0$ is higher than $\kappa_L$ and for bottom quark (right panel) $\kappa_L$ and $\kappa_T$ fall on top of each other.

\begin{figure*}[!hbt]
\begin{center}
\includegraphics[scale=0.5]{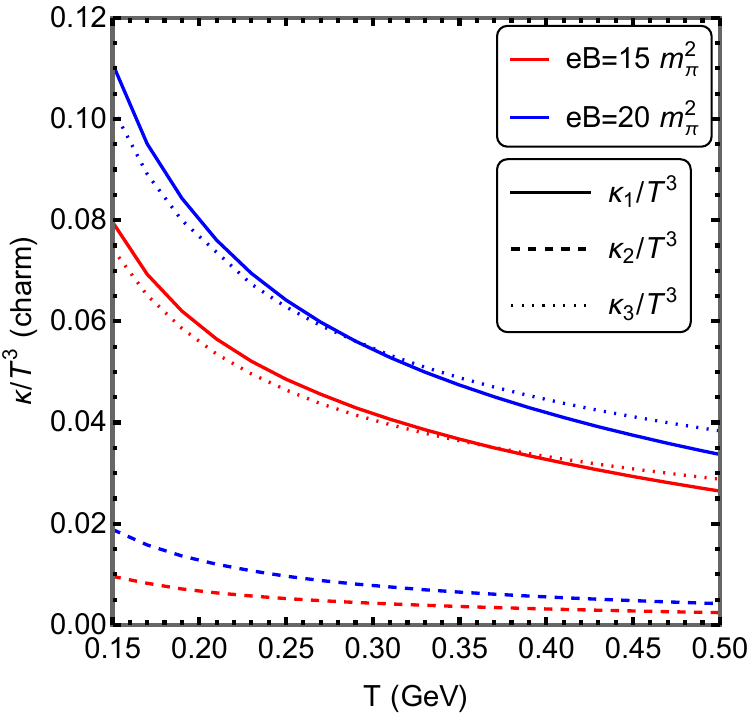}\hspace{1cm}
\includegraphics[scale=0.5]{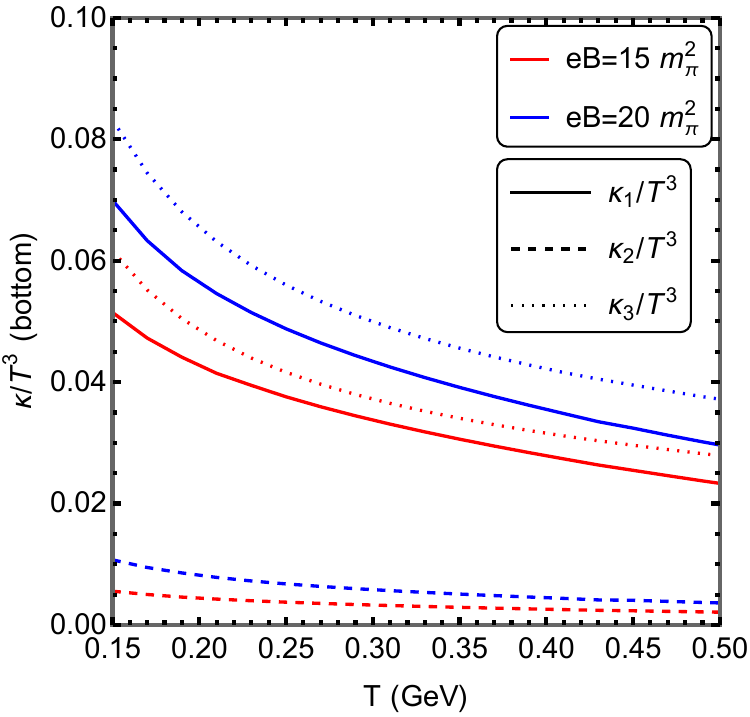}
\caption{Variation of the scaled charm (left panel) and bottom (right panel) quark momentum diffusion coefficients (for $\vec{v} \perp \vec{B}$) with temperature for two different values of external magnetic fields, i.e. $eB=15 m_\pi^2$ and $20m_\pi^2$. For both the cases we have shown the plots for scaled transverse components $\kappa_1$ (solid lines), $\kappa_2$ (dashed lines) and longitudinal component $\kappa_3$ (dotted lines). Charm and bottom quark masses $M$ are specified in the text and HQ momentum $p$ is taken to be 1 GeV.  } 
\label{kappavT_scaled_cb_vperpB}
\end{center}
\end{figure*}


\begin{figure*}[!hbt]
\begin{center}
\includegraphics[scale=0.55]{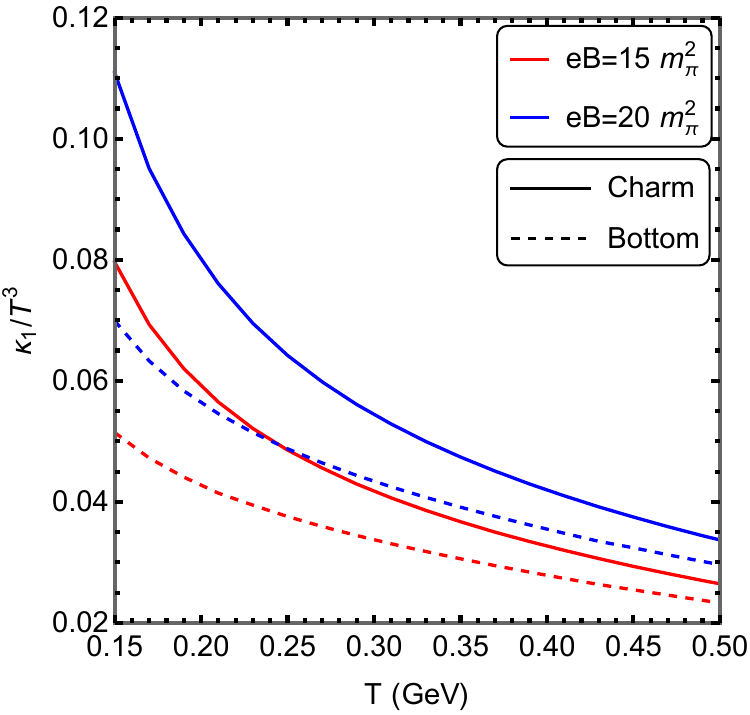}
\includegraphics[scale=0.55]{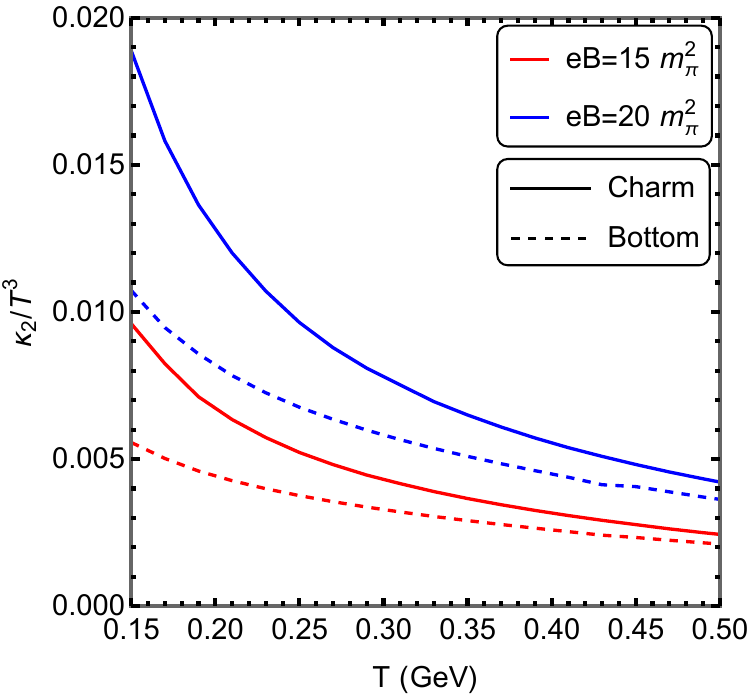}
\includegraphics[scale=0.55]{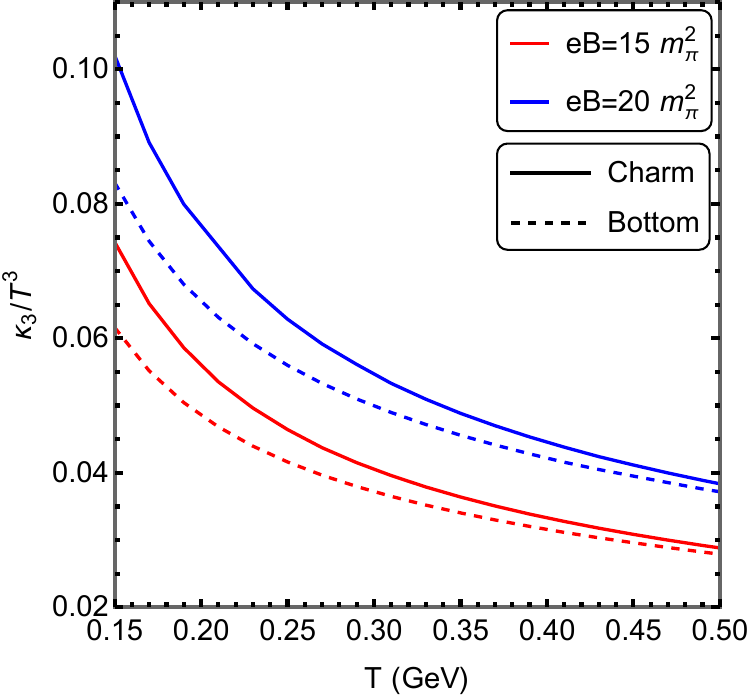}
\caption{Variation of the scaled HQ transverse ($\kappa_1$ and $\kappa_2$, top 2 panels) and longitudinal ($\kappa_3$, bottom panel) momentum diffusion coefficients (for $\vec{v} \perp \vec{B}$) with temperature for two different values of external magnetic fields, i.e. $eB=15m_\pi^2$ and $20m_\pi^2$. In each plot, we have shown the variations for charm (solid lines) and bottom (dashed lines) quarks. Heavy quark masses $M$ are specified in the text and momentum $p$ is taken to be 1 GeV. }
\label{kappavT_scaled_123_vperpB}
\end{center}
\end{figure*}

\begin{figure*}[!hbt]
\begin{center}
\includegraphics[scale=0.55]{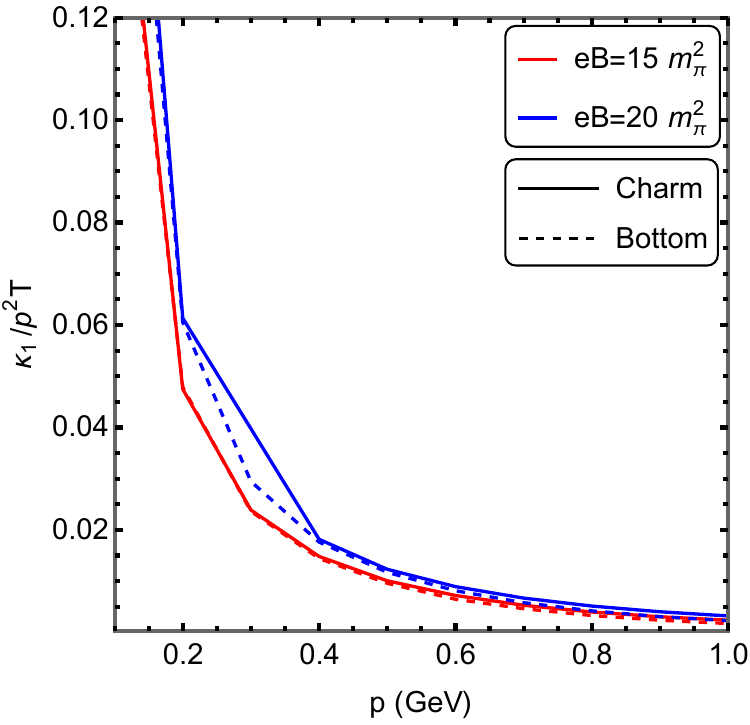}
\includegraphics[scale=0.55]{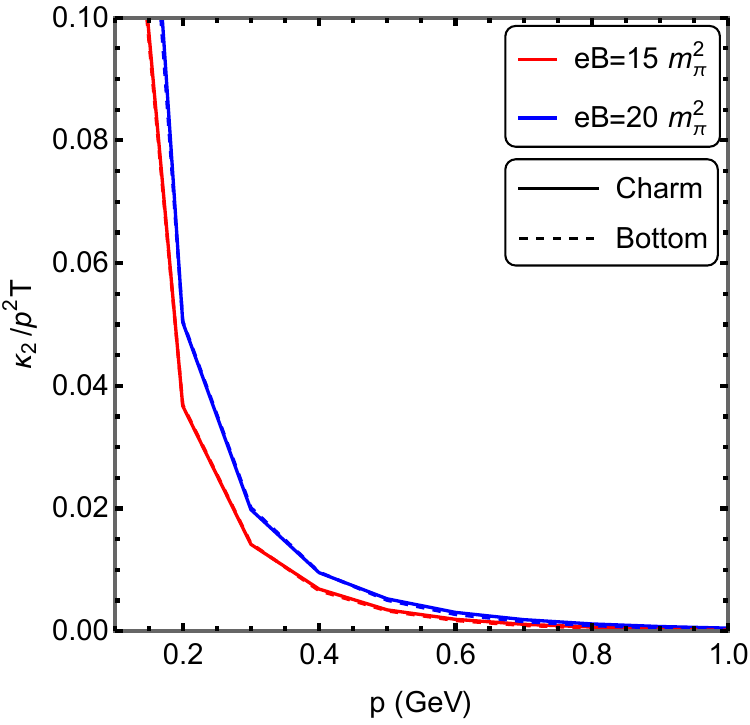}
\includegraphics[scale=0.55]{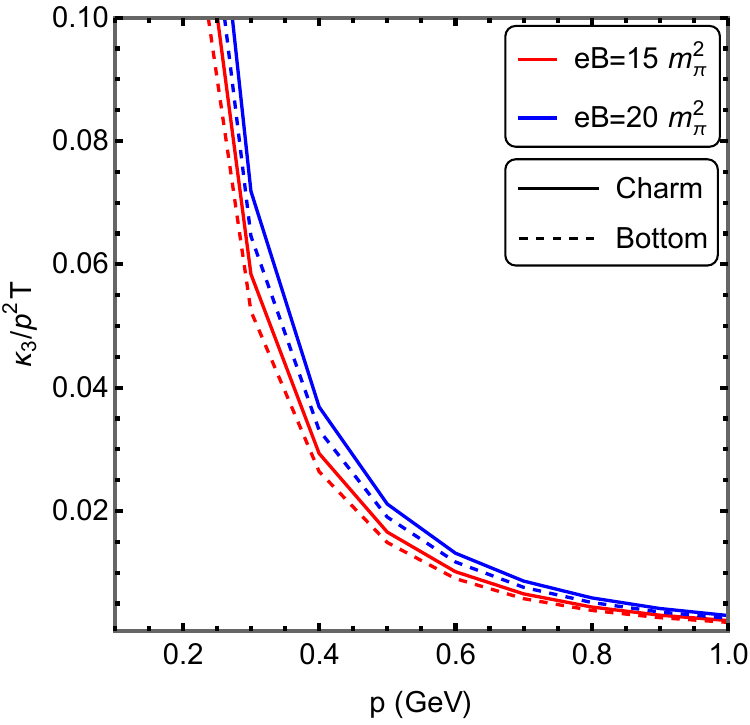}
\caption{Dependence of the HQ  transverse (top two panels) and longitudinal (bottom panel) momentum diffusion coefficients (for $\vec{v} \perp \vec{B}$), normalized by $p^2 T$, on the HQ momentum $p$ for two different values of external magnetic fields, i.e. $eB=15m_\pi^2$ and $20m_\pi^2$. In each plot we have presented curves for both charm (solid lines) and bottom (dashed) quarks. Heavy quark masses $M$ are specified in the text and the temperature $T$ is taken to be 0.2 GeV.  } 
\label{kappavp_vperpB}
\end{center}
\end{figure*}

We have also shown the variation of $\kappa_T$ with temperature for charm quark with two different values of the external momentum $p$ in Fig. \ref{kappaT_scaled_vp}, i.e. $p=1$ GeV and $p=0.5$ GeV. Again we have chosen two different values of the magnetic field, $eB= 15m_\pi^2$ and $20m_\pi^2$. This plot is done to check the consistency of our calculation as we have maintained the scale hierarchy of $M\gg p$ ( $M$ is the heavy quark mass) and simplified our expressions accordingly. For bottom quark mass $M=4.18$ GeV this condition is easily satisfied. But for charm quark mass, since $M=1.28$ GeV, and we have chosen $p=1$ GeV for most of our results, it was necessary to compare with a different (smaller) value of $p$. It can be seen from figure \ref{kappaT_scaled_vp} that the behavior for two different values of $p$ are almost identical. At all values of temperature the $\kappa_T$ is bigger at larger HQ momentum  for both values of the magnetic field, i.e. $eB=15 m_\pi^2$ and $eB=20 m_\pi^2$.

\subsection{case 2 : $\vec{v} \perp \vec{B}$}
\label{case2_results}

For the $\vec{v} \perp \vec{B}$ case we have two anisotropic directions given by $\vec{v}$ and $\vec{B}$. These subsequently give rise to three different momentum coefficients, which we have noted as $\kappa_1$, $\kappa_2$ and $\kappa_3$ in the present study, representating the longitudinal ($\kappa_3$) and transverse ($\kappa_1, \kappa_2$) components. In this case the heavy quark momenta can be nonvanishing in any of the directions transverse to $\vec{B}$ direction ($z$), i.e. $x$ and/or $y$. In the following we choose a particular system where the heavy quark is chosen to be moving along the $x$ direction. Hence the heavy quark momentum has only one nonvanishing component along the $x$ direction. We discuss our findings for $\kappa_1$, $\kappa_2$ and $\kappa_3$ for charm and bottom quarks (mass $M=1.28$ GeV and  $M=4.18$ GeV respectively) moving perpendicular ($x$ direction) to an external magnetic field along the $z$ direction.

In Fig. \ref{kappavT_scaled_cb_vperpB} we have shown the variation of the scaled heavy quark momentum diffusion coefficients with temperature for two different values of external magnetic fields, i.e. $eB=15 m_\pi^2$ and $20m_\pi^2$. We have presented two separate plots for the charm (left panel) and bottom (right panel) quarks. For both the cases we have shown the variations for scaled transverse components $\kappa_1$ (solid lines), $\kappa_2$ (dashed lines) and longitudinal component $\kappa_3$ (dotted lines). One can observe from the plots that for bottom quarks, values of the longitudinal component $\kappa_3$ (dotted lines) are the largest, followed by the transverse component $\kappa_1$ (solid lines). For charm quarks, we notice a crossover between $\kappa_1$ and $\kappa_3$, where $\kappa_1$ dominates at lower $T$ and $\kappa_3$ at higher $T$. For both the plots, values of $\kappa_2$ (dashed lines), which is basically transverse to both the magnetic field and the velocity directions, appear to be the lowest of the plot, almost an order of magnitude lower than $\kappa_1 /\kappa_3$. Also we can see that with an increasing magnetic field, values for all the HQ momentum diffusion components have also increased.

Fig. \ref{kappavT_scaled_123_vperpB} shows the similar variation as in Fig. \ref{kappavT_scaled_cb_vperpB}, but this time the representation is different. Here we have compared charm (solid lines) and bottom (dashed lines) quark curves together for three different plots, one each for $\kappa_1$ (top left panel), $\kappa_2$ (top right panel) and $\kappa_3$ (bottom panel). 
For all three components, $\kappa_1$, $\kappa_2$ and $\kappa_3$, the charm quark momentum diffusion coefficients are found to be considerably larger than that of the bottom quark, especially at relatively lower temperature region.

Finally in Fig. \ref{kappavp_vperpB} we have shown the dependence of the  transverse (top two panels) and longitudinal (bottom panel) momentum diffusion coefficients on HQ momentum $p$ for two different values of external magnetic fields, i.e. $eB=15m_\pi^2$ and $20m_\pi^2$. In each plot we have presented curves for both charm (solid lines) and bottom (dashed) quarks. The temperature in these plots is taken to be $T=0.2$ GeV. 
Note that these $\kappa$ coefficients characterize the momentum-squared transfer due to medium kicks, therefore it would be more meaningful to examine a dimensionless combination $\frac{\kappa}{p^2 T}$. This ratio is constructed with the following thinking: the $\kappa$ multiplying the medium time scale $1/T$ gives the average change in momentum-squared  $\langle \delta p^2 \rangle $ due to medium kicks over that time scale, which is to be compared with the original momentum square $p^2$ of the particle.
The plots for the transverse momentum diffusion coefficients $\kappa_1$ and $\kappa_2$ suggest that at lower values of HQ momentum, bottom and charm quark transverse momentum diffusion coefficients are almost equal while for higher values of HQ momentum the charm quark transverse momentum diffusion coefficients become larger than  the bottom quark. For the longitudinal coefficient $\kappa_3$ the charm quark momentum diffusion coefficients are always visibly larger than that of   the bottom quark. 
The results show a monotonic decrease with increasing HQ momentum, suggesting a reduced influence of medium kicks for heavy quarks with larger momenta. The results for $eB=15 m_\pi^2$ and $eB=20 m_\pi^2$ are fairly close, while both being considerably smaller as compared with the zero magnetic field case. Such a behavior may be related to the lowest Landau level approximation which reduces the available scattering states of the medium quarks. Phenomenologically, this may suggest a suppression of the heavy quark diffusion at the very early stage of the QGP evolution when the magnetic field is very strong. With future quantitative simulations of heavy quark transport with magnetic-field-dependent diffusion coefficients, one could hope for putting constraints on the lifetime of magnetic field in these collisions.


\section{Summary}
\label{sec6}

In summary,   we have studied the momentum diffusion coefficients for heavy quarks (charm and bottom) moving in a hot quark-gluon plasma under the presence of a strong external magnetic field along the $z$ direction. We have considered two specific cases, i.e. when the HQ is moving parallel to the external magnetic field ($\vec{v}\sp\vec{B}$) and when the HQ is moving perpendicular to the external magnetic field ($\vec{v}\perp\vec{B}$). For these two cases we have evaluated the relevant momentum diffusion coefficients within the HTL approximation. To incorporate the soft gluonic momenta in our evaluation, we have worked with the recently obtained effective HTL gluon propagator in a hot and magnetized medium~\cite{Karmakar:2018aig}. For $\vec{v}\sp\vec{B}$, we have one anisotropic direction along $z$ which results in two different momentum diffusion coefficients, longitudinal $\kappa_L$ and transverse $\kappa_T$. On the other hand for $\vec{v}\perp\vec{B}$ we have two different anisotropic direction (in our case we have chosen that the HQ is moving along $x$ direction) which results in three different momentum diffusion coefficients along three spatial directions, i.e. $\kappa_1, \kappa_2$ and $\kappa_3$. Considering the $\vec{B}$ direction as our reference, we have called $\kappa_3$ as the longitudinal and $\kappa_{1,2}$ as two transverse coefficients.  For all these different $\kappa$'s, we have shown the variation with temperature for different values of $eB$, both for charm and bottom quarks which revealed some interesting features. Many of these results are obtained for the first time. Numerical evaluations demonstrate a considerable influence of the strong magnetic field on these coefficients for $eB$ values accessible in high energy heavy ion collisions. It may be noted that the present calculations can be  adapted to numerically evaluate the fully anisotropic drag coefficients for the HQ velocity in arbitrary direction. 
In the present study we focus on showing results for the momentum diffusion $\kappa$ coefficients and it shall be noted that the corresponding $\eta$ drag coefficients can be directly obtained via their relations to the $\kappa$ coefficients as in Eq.~(\ref{eq_eta_kappa}).

A natural next step is to go beyond the LLL approximation adopted in the present work under the assumption of extremely strong magnetic field. This is a very challenging task but may be important for realistic applications. It would also be highly interesting to explore the phenomenological implications  of our theoretical results. For example, one could implement the $eB$ and HQ $\vec{v}$ dependent drag coefficients into a Langevin transport code (e.g.~\cite{Li:2019lex}) and examine the dynamical HQ in-medium evolution. In particular, there could be nontrivial consequence of the anisotropic transport coefficients due to the magnetic field for experimental observables such as directed and elliptic flow of the open heavy flavor mesons. We expect to report progress along these lines in a future work.

\begin{acknowledgments}
This work is supported in part by the Guangdong Major Project of Basic and Applied Basic Research No. 2020B0301030008, Science and Technology Program of Guangzhou Project No. 2019050001, the National Natural Science Foundation of China under Grant No. 12022512, No. 12035007, No. 11735007, as well as by the NSF Grant No. PHY-1913729.
\end{acknowledgments}

\appendix
\section{General Structure of an effective gauge boson propagator in a magnetized medium}
\label{appA}
We begin this section by defining Lorentz scalars, vectors and tensors that characterize 
the heat bath or hot medium in a local rest frame:
\begin{align}
u^\mu &=(1,0,0,0), \\
Q^\mu u_\mu&=Q\cdot u=q_0.\nn
\end{align}
In the rest frame of the heat bath, another anisotropic four-vector $n^\mu$ can  be 
defined uniquely as projection of the EM field tensor $F^{\mn}$ along $u^\mu$,
\be
 n_\mu \equiv \frac{1}{2B} \epsilon_{\mu\nu\rho\lambda}\, u^\nu F^{\rho\lambda} 
 = \frac{1}{B}u^\nu {\tilde F}_{\mu\nu} = (0,0,0,1), \label{bmu}
 \ee
which represents the $z$-direction. This also establishes a connection between the heat bath and the magnetic field. 

We first form the transverse four momentum and the transverse metric tensor as  
\begin{subequations}
\begin{align}
\tilde{Q}^\mu &= Q^\mu - (Q\cdot u)u^\mu, \\
Q_\perp^\mu &=\tilde{Q}^\mu + (Q\cdot n)n^\mu \nn\\
&= Q^\mu - q_0 u^\mu + q^3 n^\mu = Q^\mu-Q_\sp^\mu,\\
\tilde{g}^{\mn} &= g^{\mu\nu} - u^\mu u^\nu ,\\
g_\perp^{\mn} &= \tilde{g}^{\mn} + n^\mu n^\nu = g^{\mu\nu} - g_\sp^{\mu\nu},
\end{align}
\end{subequations}
where 
\begin{subequations}
\begin{align}
Q_\sp^\mu &=  q_0 u^\mu -q^3 n^\mu, \label{p_sp}\\
Q_\sp^2 &= Q_\sp^\mu Q^\sp_\mu = q_0^2-q_3^2,\\
g_\sp^{\mu\nu} &= u^\mu u^\nu - n^\mu n^\nu,\label{eta_sp}\\
Q_\perp^\mu Q^\perp_{\mu} &=Q_\perp^2= Q^2-q_0^2+q_3^2=Q^2-Q_\sp^2=-q_\perp^2,
\end{align}
\end{subequations}
where $Q^2=Q_\sp^2+Q_\perp^2=Q_\sp^2-q_\perp^2$, $Q_\sp^2=q_0^2-q_3^2$ and $q_\perp^2=q_1^2+q_2^2$. 
We further note  that the three independent Lorentz scalars are  $q_0$, $q^3=Q\cdot n$  and $Q_\perp^2$.
One can further redefine four vector $u^\mu$ as
\bea
\bar{u}^\mu = u^\mu - \frac{(Q\cdot u)Q^\mu}{Q^2} 
= u^\mu - \frac{q_0 Q^\mu}{Q^2}.
\eea
which is orthogonal to $Q^\mu$ and similarly $n^\mu$ as
\bea
\bar{n}^\mu = n^\mu - \frac{(\tilde{Q}\cdot n)\tilde{Q}^\mu}{\tilde{Q}^2} 
= n^\mu - \frac{q_3 Q^\mu}{q^2}+ \frac{q_0q_3 u^\mu}{q^2},
\eea
which is orthogonal to $\tilde{Q}^\mu$. Now three independent and mutually transverse second rank projection tensors can be constructed in terms of  those redefined set of four-vectors and tensors as
 \begin{subequations}
\begin{align}
\Delta_1^{\mu\nu} &= \frac{\bar{u}^\mu\bar{u}^\nu}{\bar{u}^2} ,\\
\Delta_2^{\mn} &=g_{\perp}^{\mn}-\frac{Q^{\mu}_{\perp}Q^{\nu}_{\perp}}{Q_{\perp}^2},\\
\Delta_3^{\mu\nu} &=  \frac{{\bar n}^\mu {\bar n}^\nu}{\bar n^2} .
\end{align}
\end{subequations}

Next one can construct the fourth tensor as
\bea
\Delta_4^{\mn}&=& \frac{\bar u^{\mu}\bar n^{\nu}+\bar u^{\nu}\bar n^{\mu}}{\sqrt{\bar u^2}\sqrt{\bar n^2}},
\eea
which satisfies the following properties
\begin{subequations}
\begin{align}
(\Delta_4)^{\mu\rho}(\Delta_4)_{\rho \nu}&= (\Delta_1)^{\mu}_{\nu}+(\Delta_3)^{\mu}_{\nu},\\
(\Delta_k)^{\mu\rho}(\Delta_4)_{\rho\nu}&+(\Delta_4)^{\mu\rho}(\Delta_k)_{\rho\nu}=(\Delta_4)^{\mu}_{\nu},\\
(\Delta_2)^{\mu\rho}(\Delta_4)_{\rho\nu}&=(\Delta_4)^{\mu\rho}(\Delta_2)_{\rho\nu}=0.
\end{align}
\end{subequations}
with $(k=1,3)$. Now, one can write a general covariant structure of gauge boson self-energy
as 
\bea
\Pi^{\mn} = \sum\limits_{i=1}^4 d_i\Delta_i^{\mn}, \label{gen_pimn}
\eea
where $d_i$'s are four Lorentz-invariant form factors associated with
the four basis tensors given in Eqs.~(\ref{ff_d1}-\ref{ff_d4}).

The inverse of the effective gauge boson propagator can be expressed in terms of the Dyson Schwinger equation as,
\bea
\mathcal{D}_{\mn}^{-1} = (\mathcal{D}_0)^{-1}_{\mn} - \Pi_{\mn},\label{dse}
\eea
where $(\mathcal{D}_0)_{\mn}$ is the gauge boson propagator in vacuum. Using Eq.~(\ref{gen_pimn}), Eq.~(\ref{dse}) and the fact that $(\mathcal{D}^{\mr})^{-1}\mathcal{D}_{\rn}=g^\mu_\nu$ one can write down the general covariant structure of the gauge boson propagator in covariant gauge as expressed in Eq.~(\ref{gauge_prop}).

\section{Form Factors within LLL approximation}
\label{appB}

The fermion propagator within LLL approximation is given in Eq.~(\ref{prop_sfa}). Using that propagator, the fermionic contribution of the gluon self energy was computed in Ref~\cite{Karmakar:2018aig} and given as 
\begin{align}
\Pi_{\mu\nu}^s(Q) &= ~-\sum_fe^{{-q_\perp^2}/{2|q_fB|}}~\frac{g^2 |q_fB|}{2\pi}\nn\\ &T\sum\limits_{k_0}\int\frac{dk_3}{2\pi} \frac{{\cal S}_{\mu\nu}^s}{(K_\shortparallel^2-m_f^2)(R_\shortparallel^2-m_f^2)}, 
\label{Pi_mn_s}
\end{align}
with $Q$ is the external gluon momentum, $K$ is the fermion loop momentum and $R\equiv K-Q$. The tensor structure ${\cal S}_{\mu\nu}^s$ originates from the Dirac trace and given as
\begin{align}
{\cal S}_{\mu\nu}^s &= K_\mu^\shortparallel R_\nu^\shortparallel + R_\mu^\shortparallel K_\nu^\shortparallel 
- g_{\mu\nu}^\shortparallel \left((K\cdot R)_\shortparallel -m_f^2\right) \nn\\
&= u_\mu u_\nu  \left( k_0 r_0 + k_3 r_3 +m_f^2\right) + n_\mu n_\nu  \left( k_0 r_0 + k_3 r_3 -m_f^2\right)\nn\\
& - \left( u_\mu n_\nu + n_\mu u_\nu \right) \left( k_0 r_3 + k_3 r_0 \right).
\label{se_sfa_gen}
\end{align}
On the other hand the Yang-Mills(YM) contribution of the gluon self energy from the ghost and gluon loop is depicted as $\Pi_{\mn}^{g}$, which remains unaffected in presence of magnetic field and can be written as
\bea
\Pi_{\mn}^{g}(Q)=-\frac{N_cg^2T^2}{3} \int \frac{d \Omega}{2 \pi}\left(\frac{q_{0} \hat{K}_{\mu} \hat{K}_{\nu}}{\hat{K} \cdot Q}-g_{\mu 0} g_{\nu 0}\right),
\label{Pi_mn_g}
\eea
and $\mathcal{T}_Q$ is defined as 
\begin{align}
&\mathcal{T}_Q = \frac{q_0}{2q}\ln\left(\frac{q_0+q}{q_0-q} \right).
\end{align}
The total gluon self energy is then given by $\Pi_{\mn} = \Pi_{\mn}^s +\Pi_{\mn}^g$.

Now we can evaluate the form factors in Eqs.~(\ref{ff_d1}),~(\ref{ff_d2}) ,~(\ref{ff_d3}) and ~(\ref{ff_d4}) in strong field approximation as 
\begin{align}
d_1 &= \Delta_1^{\mn}(\Pi_{\mn}^{g}+\Pi_{\mn}^s)=d_1^{YM}+d_1^s
\end{align}
where
\begin{align}
d_1^{YM}&=\frac{C_Ag^2T^2}{3\bar u^2}\left[1-\mathcal{T}_Q(q_0,q)\right],
\end{align}
and 
\begin{align}
&d_1^s= -\sum_fe^{{-q_\perp^2}/{2|q_fB|}}~\frac{g^2 |q_fB|}{2\pi 
\bar{u}^2}\nn\\
& \times T \sum\limits_{k_0}\int\frac{dk_3}{2\pi} 
\frac{k_0 r_0 + k_3r_3 
+m_f^2}{(K_\shortparallel^2-m_f^2)(R_\shortparallel^2-m_f^2)}. \label{coeff_d1s}
\end{align}

As is usually done in  Hard Thermal Loop (HTL) calculations~\cite{Braaten:1989mz}, one assumes the external momenta to be soft and small compared with the hard momenta in the loop and uses the approximation $k_0\approx r_0$ and $k_3\approx r_3$ in the last numerator, thus obtaining:
\begin{widetext}
\begin{align}
d_1^s
\approx& -\sum_fe^{{-q_\perp^2}/{2|q_fB|}}~\frac{g^2|q_fB|}{2\pi 
	\bar{u}^2}~T\sum\limits_{k_0}\int\frac{dk_3}{2\pi} \left[\frac{1}{(K_\shortparallel^2-m_f^2)}+\frac{2\left(k_3^2+m_f^2\right)}
	{(K_\shortparallel^2-m_f^2)(R_\shortparallel^2-m_f^2)}\right] \nn\\
=& \sum_fe^{{-q_\perp^2}/{2|q_fB|}}~\frac{g^2|q_fB|}{2\pi 
	\bar{u}^2}~\int\frac{dk_3}{2\pi} \Bigg[-\frac{n_F(E_{k_3})}{E_{k_3}}\nn\\
& +\left\{\frac{n_F(E_{k_3})}{E_{k_3}}+\frac{q_3 k_3}{E_{k_3}}\frac{\partial n_F(E_{k_3})}{\partial k_3}\left(\frac{q_3 k_3/E_{k_3}}{q_0^2-q_3^2(k_3/E_{k_3})^2}\right)\right\}\Bigg] \nn\\
=& \sum_fe^{{-q_\perp^2}/{2|q_fB|}}~\frac{g^2|q_fB|}{2\pi 
	\bar{u}^2}\int\frac{dk_3}{2\pi}\,\frac{q_3 k_3}{E_{k_3}} \frac{\partial n_F(E_{k_3})}{\partial E_{k_3}}\left(\frac{q_3 k_3/E_{k_3}}{q_0^2-q_3^2(k_3/E_{k_3})^2}\right) . \label{d1_m_sf}
\end{align}
\end{widetext}
Using Eqs.~(\ref{coeff_d1s}) and (\ref{d1_m_sf}) one also can directly calculate the Debye screening mass in a strongly magnetized hot medium within QCD as
\begin{align}
m_D^2 &= \left.{\bar u}^2 d_1 \right |_{q_0=0, \vec{q} \rightarrow 0}=(m_D^2)_g+\sum_f \delta m_{D,f}^2
\end{align}
where $(m_D^2)_g = \frac{g^2N_c T^2}{3}$ and
\begin{align}
 \delta m_{D,f}^2 &= \frac{g^2|q_fB|}{2\pi  T} \int\limits_{-\infty}^\infty\frac{dk_3}{2\pi}~ n_F(E_{k_3})\left(1-n_F(E_{k_3})\right),
 \label{dby_m_sf}
\end{align}
which matches with the well-known expressions of QED Debye mass~\cite{Alexandre:2000jc,Bandyopadhyay:2016fyd} without the QCD factors. Now using Eq.~(\ref{dby_m_sf}) in Eq.~(\ref{d1_m_sf}) along with $E_{k_3}\sim k_3$, the form factor $d_1$ can be finally expressed  in terms of $\delta m_D$ as
\begin{align}
 d_1 &=\frac{C_Ag^2T^2}{3\bar u^2}\left[1-\mathcal{T}_Q(q_0,q)\right]\nn\\
 &-\sum_f e^{{-q_\perp^2}/{2 |q_fB|}}~\left(\frac{\delta m_{D,f}}{\bar u}\right)^2\frac{q_3^2}{q_0^2-q_3^2}
 \label{coeff_d1_final}.
\end{align}

For the form factor $d_2$, the fermionic loop doesn't contribute, and it yields
\begin{align}
d_2 &= \Delta_2^{\mn}(\Pi_{\mn}^{g}+\Pi_{\mn}^s)=d_2^{YM}+0 \nn\\
&=\frac{C_Ag^2T^2}{3}\frac{1}{2}\left[\frac{q_0^2}{q^2}-\frac{Q^2}{q^2}\mathcal{T}_Q(q_0,q)\right].
 \label{coeff_d2}
\end{align}

For the form factor $d_3$, we apply the similar procedure as done for $d_1$ and one obtains
\begin{align}
d_3 =& \Delta_3^{\mn}(\Pi_{\mn}^{g}+\Pi_{\mn}^s)= d_3^{YM}+d_3^s\nn\\
=&\frac{C_Ag^2T^2}{3}\frac{1}{2}\left[\frac{q_0^2}{q^2}-\frac{Q^2}{q^2}\mathcal{T}_Q(q_0,q)\right]+\sum_f e^{{-q_\perp^2}/{2  |q_fB|}}\nn\\
&\times\frac{g^2|q_f B|}{2\pi}\frac{q_{\perp}^2}{q^2}~ T \sum_{k_0}\int\frac{dk_3}{2\pi} \frac{k_0 r_0+k_3r_3-m_f^2}{(K_\sp^2-m_f^2)(R_\sp^2-m_f^2)} \nn\\
\approx& \frac{C_Ag^2T^2}{3}\frac{1}{2}\left[\frac{q_0^2}{q^2}-\frac{Q^2}{q^2}\mathcal{T}_Q(q_0,q)\right]\nn\\
&+\sum_f e^{{-q_\perp^2}/{2 |q_fB|}}~\delta m_{D,f}^2~\frac{q_{\perp}^2}{q^2}\frac{q_3^2}{q_0^2-q_3^2}.
\label{coeff_d3}
\end{align}

Finally for the last form factor $d_4$ the YM contrbution vanishes and it can be obtained as
\begin{align}
d_4 &=\frac{1}{2}\Delta_4^{\mn}(\Pi_{\mn}^{g}+\Pi^s_{\mn})=\frac{1}{2}\Delta_4^{\mn}\Pi^s_{\mn}=d_4^s,\label{coeff_d4}
\end{align}
where $d_4^s$ is given by
\begin{widetext}
\begin{align}
d_4^s=&\frac{1}{2}\Delta_4^{\mn} \Pi_{\mn}^s=\sum_fi~e^{{-q_\perp^2}/{2  |q_fB|}}~\frac{g^2|q_f B|}{4\pi \sqrt{\bar u^2}\sqrt{\bar n^2}}\int\frac{d^2K_\sp}{(2\pi)^2} \bigg[\frac{-2\frac{\bar u \cdot n}{\bar u^2}\big(k_0^2+k_3^2+m_f^2\big)+4k_0k_3}{(K_\sp^2-m_f^2)(R_\sp^2-m_f^2)}\bigg] \nn\\
=&\sum_f e^{{-q_\perp^2}/{2  |q_fB|}}~\frac{g^2|q_f B|}{4\pi \sqrt{\bar u^2}\sqrt{\bar n^2}}\int\frac{dk_3}{2\pi} \nn\\ 
&\times\bigg[-2\frac{\bar u \cdot n}{\bar u^2} \frac{\partial n_F(E_{k_3})}{\partial E_{k_3}}\frac{q_3^2 k_3^2/E_{k_3}^2}{\big(q_0^2-q_3^2 k_3^2/E_{k_3}^2\big)}+\frac{2\partial n_F(E_{k_3})}{\partial E_{k_3}}\frac{q_0q_3 k_3^2/E_{k_3}^2}{\big(q_0^2-q_3^2 k_3^2/E_{k_3}^2\big)}\bigg] \nn\\
\approx& \sum_f ~e^{{-q_\perp^2}/{2  |q_fB|}} \frac{\sqrt{\bar n^2}}{\sqrt{\bar u^2}}\delta m_{D,f}^2~ \frac{q_0q_3}{q_0^2-q_3^2},\label{d4_m_sf}
\end{align}
\end{widetext}
where   $\bar n^2=-q_{\perp}^2/q^2$ and $\bar u^2=-q^2/Q^2$.

\section{Spectral functions $\rho_i$'s}
\label{appC}

The explicit expressions for the spectral functions are given by,
\begin{align}
\rho_1(\omega,q) =& -\frac{1}{\pi} {\rm Im}\left(\mathcal{J}_1\Big |_{q_0 = \omega +i\epsilon}\right) \nn\\
=& -\frac{1}{\pi} {\rm Im}\left(\frac{(Q^2-d_3)}{(Q^2-d_1)(Q^2-d_3)-d_4^2} \Bigg |_{q_0 = \omega +i\epsilon}\right) \nn\\
=& -\frac{1}{\pi D}\Bigl[\Im_{d_1}\left(\Im_{d_3}^2+\Re_{d_3}^2+Q^4-2Q^2\Re_{d_3} \right)  \nn\\
&+2\Im_{d_4}\Re_{d_4}\left(Q^2-\Re_{d_3}\right) +\Im_{d_3}\left(\Re_{d_4}^2-\Im_{d_4}^2\right)\Bigr].
\end{align}

Here $\Im_{d_i}$ and $\Re_{d_i}$ respectively depict the imaginary and real parts of $d_i$'s.

\begin{align}
\rho_2(\omega,q) =& -\frac{1}{\pi} {\rm Im}\left(\mathcal{J}_2\Big |_{q_0 = \omega +i\epsilon}\right) \nn\\
=& -\frac{1}{\pi} {\rm Im}\left(\frac{1}{(Q^2-d_2)}  \Bigg |_{q_0 = \omega +i\epsilon}\right) \nn\\
=& -\frac{1}{\pi}\left[\frac{\Im_{d_2}}{\Im_{d_2}^2-(Q^2-\Re_{d_2})^2}\right], \\
\rho_3(\omega,q) =& -\frac{1}{\pi} {\rm Im}\left(\mathcal{J}_3\Big |_{q_0 = \omega +i\epsilon}\right) \nn\\
=& -\frac{1}{\pi} {\rm Im}\left(\frac{(Q^2-d_1)}{(Q^2-d_1)(Q^2-d_3)-d_4^2} \Bigg |_{q_0 = \omega +i\epsilon}\right) \nn\\
=& -\frac{1}{\pi D}\Bigl[\Im_{d_3}\left(\Im_{d_1}^2+\Re_{d_1}^2+Q^4-2Q^2\Re_{d_1} \right) \nn\\
&+ 2\Im_{d_4}\Re_{d_4}\left(Q^2-\Re_{d_1}\right) +\Im_{d_1}\left(\Re_{d_4}^2-\Im_{d_4}^2\right)\Bigr],\\
\rho_4(\omega,q) =& -\frac{1}{\pi} {\rm Im}\left(\mathcal{J}_4\Big |_{q_0 = \omega +i\epsilon}\right) \nn\\
=& -\frac{1}{\pi} {\rm Im}\left(\frac{d_4}{(Q^2-d_1)(Q^2-d_3)-d_4^2} \Bigg |_{q_0 = \omega +i\epsilon}\right) \nn\\
=& -\frac{1}{\pi D}\Bigl[\Im_{d_4}\Bigl(-\Im_{d_1}\Im_{d_3}+\Re_{d_1}\Re_{d_3} \nn\\
&+\Re_{d_4}^2+\Im_{d_4}^2+Q^4-Q^2(\Re_{d_1}+\Re_{d_3}) \Bigr) \nn\\
&+ \Re_{d_4}\Bigl(Q^2(\Im_{d_1}+\Im_{d_3})\Im_{d_3}\Re_{d_1}-\Im_{d_1}\Re_{d_3}\Bigr)\Bigr].
\end{align}
Here the denominator $D$ is expressed as
\begin{align}
D =& \Bigl[ \Bigl( -\Im_{d_1}Q^2 - \Im_{d_3}Q^2 + \Im_{d_3}\Re_{d_1} + \Im_{d_1}\Re_{d_3} - 2\Im_{d_4}\Re_{d_4}\Bigr)^2 \nn\\
&+\Bigl(-\Im_{d_1}\Im_{d_3} + \Im_{d_4}^2 + (Q^2-\Re_{d_1})(Q^2-\Re_{d_3}) -\Re_{d_4}^2\Bigr)^2\Bigr]
\end{align}

Next we evaluate $\Re_{d_i}$'s and $\Im_{d_i}$'s, i.e. real and imaginary parts of $d_i$'s.                                                                                                                             The imaginary parts of $d_i$'s come from $\mathcal{T}_Q$ and the factor $\frac{q_3}{q_0^2-q_3^2}$, which subsequently can be given as follows - 
\begin{align}
\Im_{d_1} =& \frac{C_Ag^2T^2}{3\bar{u}^2}\frac{\pi \omega}{2q}+ \frac{\pi\omega}{2}\sum_f e^{{-q_\perp^2}/{2 |q_fB|}}\nn\\
&~\times \frac{\delta m_{D,f}^2}{\bar{u}^2}\left[\delta(\omega+q_3) +\delta(\omega-q_3) \right], \\
\Im_{d_2} =& \frac{C_Ag^2T^2}{3}\frac{\pi \omega Q^2}{4q^3}, \\
\Im_{d_3} =& \frac{C_Ag^2T^2}{3}\frac{\pi \omega Q^2}{4q^3}-\frac{\pi\omega}{2}\sum_f e^{{-q_\perp^2}/{2 |q_fB|}}\nn\\
&~\times \delta m_{D,f}^2~\frac{q_{\perp}^2}{q_{\perp}^2+\omega^2}\left[\delta(\omega+q_3) +\delta(\omega-q_3) \right], \\
\Im_{d_4} =& \frac{\pi\omega}{2}\sum_f e^{{-q_\perp^2}/{2 |q_fB|}}\nn\\
&\times~\delta m_{D,f}^2~\frac{\sqrt{\bar{n}^2}}{\sqrt{\bar{u}^2}}\left[\delta(\omega+q_3) -\delta(\omega-q_3) \right].
\end{align}
The real parts of $d_i$ can be expressed in the same way as Eqs.~(\ref{coeff_d1_final}), (\ref{coeff_d2}), (\ref{coeff_d3}) and (\ref{coeff_d4}), with replacing $\ln\left(\frac{q_0+q}{q_0-q} \right)$ by $\ln\left|\frac{q_0+q}{q_0-q} \right|$ within $\mathcal{T}_Q$ and by considering the principle value for the factor $\frac{q_3}{q_0^2-q_3^2}$.

\section{Discussion on the static limit}
\label{appD}

In this appendix we examine  the static limit, which means taking $\vec{v}\to 0$, from our general expression at finite velocity. 
Specifically we consider the case-1, i.e. $\vec{v}\shortparallel \vec{B}$. We start with our expression from Eq.~\ref{kappaT_case1_final} to compare with Eq.~(4.34) of Ref~\cite{Fukushima:2015wck}. In the static, i.e. $v\rightarrow 0$ limit, it can be expressed as  

\begin{align}
\kappa_T\big|_{v\rightarrow 0} &= \frac{\pi g^2T}{16E^2}\sum_{i=1}^4\int\frac{d^3q}{(2\pi)^3}q_\perp^2 e^{-q_\perp^2/|q_fB|}\nn\\
&\left[\coth(\frac{\omega}{2T})\rho_i(\omega,q) A_i^{(1)} \right]_{\omega\rightarrow 0}
\label{kappaT_initial_sl}
\end{align}

Now, evaluating the $\omega\rightarrow 0$ limits of the real and imaginary components of the spectral functions and $A_i^{(1)}$'s, we obtain that the only non-vanishing term comes from $i=1$, i.e. 
\begin{align}
&\left[\coth(\frac{\omega}{2T})\rho_1(\omega,q)A_1^{(1)}\right]_{\omega\rightarrow 0} \nn\\
&= \left((m_D^2)_g \frac{\pi T}{q} + 2\pi T\delta(q_3)s(q_\perp)\right) \frac{8E^2}{(q^2+\Re_{d_1}^0)^2}
\end{align}
with
\begin{align}
\Re_{d_1}^0 &= \Re_{d_1}\big|_{\omega\rightarrow 0} = (m_D^2)_g + s(q_\perp),
\end{align}
where $s(q_\perp) = \sum_f e^{{-q_\perp^2}/{2 |q_fB|}}~\delta m_{D,f}^2$
 and $\delta m_{D,f}^2 \approx \frac{\alpha_s|q_fB|}{\pi}$ (using Eq.~\ref{dby_m_sf}). All the other terms (for $i=2,3,4$) vanish in the static limit of $\omega \rightarrow 0$ either due to vanishing spectral functions or vanishing $A_i^{(1)}$'s. 

Combining all these we get the expression for the transverse momentum diffusion coefficient in the static limit from Eq.~\ref{kappaT_initial_sl} as, 
\begin{align}
\kappa_T\big|_{v\rightarrow 0} &= \frac{\pi g^2T}{2}\int\frac{d^3q}{(2\pi)^3}q_\perp^2 e^{-q_\perp^2/|q_fB|}\nn\\
&\left((m_D^2)_g \frac{\pi T}{q} + 2\pi T\delta(q_3)s(q_\perp)\right) \frac{1}{(q^2+\Re_{d_1}^0)^2}.
\end{align}

Now if we remove the pure glue part from our expression, we see that the transverse momentum diffusion coefficient comes out to be
\begin{align}
\kappa_T^q\big|_{v\rightarrow 0} &= \frac{ g^2T^2}{8\pi}\int d^2q_\perp ~q_\perp^2 ~e^{-q_\perp^2/|q_fB|}  \frac{s(q_\perp)}{(q^2+s(q_\perp))^2},
\end{align}
which matches with the Eq.~(4.34) of Ref~\cite{Fukushima:2015wck}.


\end{document}